%% file: main.tex
\documentclass[10pt,twocolumn,letterpaper]{article}
\usepackage{iccv}              
\usepackage[accsupp]{axessibility}
\usepackage{booktabs} 
\usepackage{siunitx} 
\usepackage{float}
\definecolor{iccvblue}{rgb}{0.21,0.49,0.74}
\usepackage[pagebackref,breaklinks,colorlinks,allcolors=iccvblue]
{hyperref}
\usepackage{wrapfig}

\def\paperID{15012} 
\def\confName{ICCV}
\def\confYear{2025}

\input{macros}

\title{\ourmethod{}: Visually-Informed and Geometry-Aware 3D Shape Abstraction}

\author{Richard Liu\\
University of Chicago\\
{\tt\small guanzhi@uchicago.edu}
\and
Daniel Fu\\
University of Chicago\\
{\tt\small danielfu@uchicago.edu}
\and
Noah Tan\\
University of Chicago\\
{\tt\small tntan@uchicago.edu}
\and
Itai Lang\\
University of Chicago\\
{\tt\small itailang@uchicago.edu}
\and
Rana Hanocka\\
University of Chicago\\
{\tt\small ranahanocka@uchicago.edu}
}

\begin{document}
\twocolumn[{%
\renewcommand\twocolumn[1][]{#1}%
\maketitle
\begin{center}
    \centering
    \captionsetup{type=figure}
    \includegraphics[width=\textwidth]{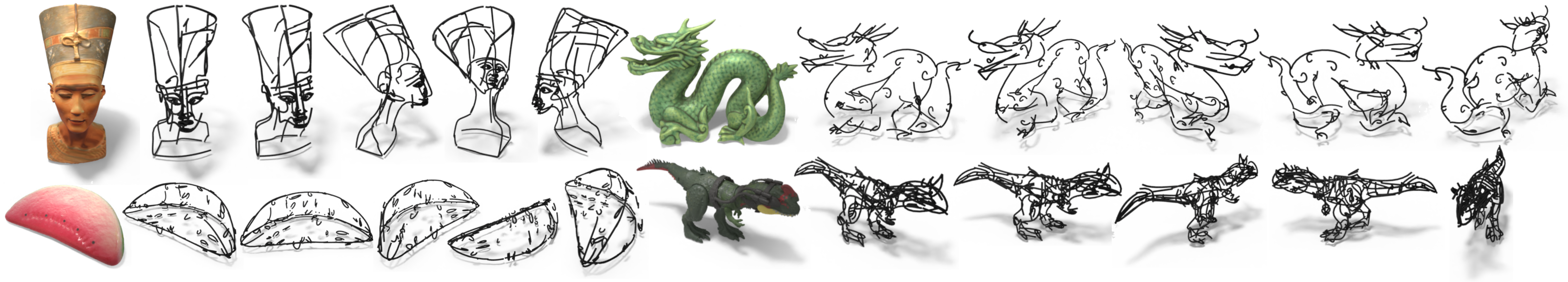}
    \captionof{figure}{\ourmethod{} produces 3D shape abstractions in the form of 3D strokes. The abstractions retain the overall shape structure and capture texture concepts (\eg dragon scales and watermelon seeds) as well as key salient features (\eg facial features).}
    \label{fig:teaser}
\end{center}
}]

\input{sec/0_abstract}

\input{sec/1_intro}
\input{sec/2_related_work}
\input{sec/3_method}

\input{sec/4_experiments}
\input{sec/5_conclusion}
\bibliographystyle{ieeenat_fullname}
\bibliography{references}
\clearpage

\input{sec/X_suppl}

\end{document}

%% file: macros.tex
\newcommand{\ourmethod}{WIR3D}

\newif\ifdraft
\draftfalse

\ifdraft
\newcommand{\rana}[1]{{\bfseries \scriptsize \color{red} Rana: #1}}
\newcommand{\itai}[1]{{\bfseries \scriptsize \color{blue} Itai: #1}}
\newcommand{\richard}[1]{{\color{teal} Richard: #1}}
\newcommand{\il}[1]{{\color{blue}#1}}

\else
\newcommand{\rana}[1]{}
\newcommand{\itai}[1]{}
\newcommand{\richard}[1]{\color{black}#1}
\newcommand{\il}[1]{{\color{black}#1}}
\fi

\makeatletter
\DeclareRobustCommand\onedot{\futurelet\@let@token\@onedot}
\def\@onedot{\ifx\@let@token.\else.\null\fi\xspace}

\def\eg{\emph{e.g}\onedot}

\def\etal{\emph{et al}\onedot}
\makeatother

\AtEndPreamble{
    \usepackage[capitalize]{cleveref}
    \crefname{section}{Sec.}{Secs.}
    \Crefname{section}{Section}{Sections}
    \Crefname{table}{Table}{Tables}
    \crefname{table}{Tab.}{Tabs.}
}

%% file: sec/0_abstract.tex
\begin{abstract}

In this work we present \ourmethod{}, a technique for abstracting 3D shapes through a sparse set of visually meaningful curves in 3D. We optimize the parameters of B\'ezier curves such that they faithfully represent both the geometry and salient visual features (e.g. texture) of the shape from arbitrary viewpoints. We leverage the intermediate activations of a pre-trained foundation model (CLIP) to guide our optimization process. 
\richard{We divide our optimization into two phases: one for capturing the coarse geometry of the shape, and the other for representing fine-grained features. Our second phase supervision is spatially guided by a novel localized keypoint loss. This spatial guidance enables user control over abstracted features. We ensure fidelity to the original surface through a neural SDF loss, which allows the curves to be used as intuitive deformation handles. We successfully apply our method for shape abstraction over a broad dataset of shapes with varying complexity, geometric structure, and texture, and demonstrate downstream applications for feature control and shape deformation.}

\end{abstract}


%% file: sec/1_intro.tex
\section{Introduction}
\label{sec:intro}
In this work, we explore whether it is possible to abstract a 3D shape into a sparse set of semantically-informed curves. A key challenge lies in finding a sparse set of curves that best represents the shape's \textit{visual features}. We use this term deliberately, to encompass both the geometry and texture features which are salient to humans. This problem cannot be solved with surface analysis alone, which focuses on low-level geometric contours. Our task is much broader, in that we wish to capture high-level concepts, visually salient geometry, and textures (e.g. \cref{fig:teaser}). 

\begin{figure*}
    \centering
    \includegraphics[width=0.85\linewidth]{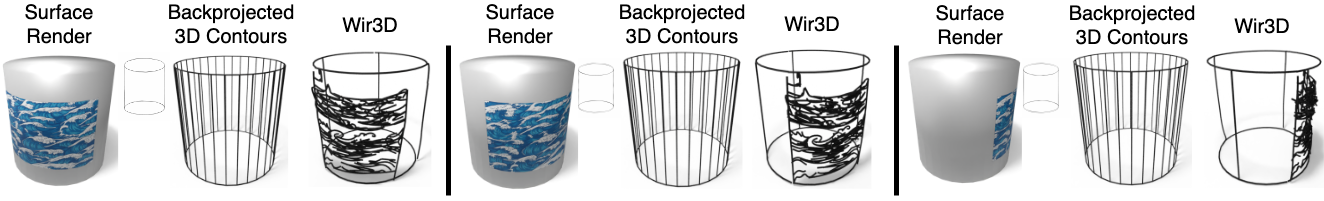}
    \caption{\textbf{Motivating example.} Abstracting even a simple shape like a cylinder is nontrivial. Rendering the contours of the cylinder (inset) results in a static image from every view, removing any sense of 3D volume. Backprojecting these contours into 3D results in a dense cluster of lines spanning the body of the cylinder, which is both non-sparse and unsatisfying aesthetically. In contrast, our method effectively abstracts the 3D geometry of the cylinder with a few strokes along with the texture.}
    \label{fig:cylinder-ex}
\end{figure*}

Existing works have dealt with the problem of computing occluding contours for non-photorealistic rendering \cite{liu2021neural, capouellez2023algebraic, liu2023contesse}. Occluding contours relies entirely on surface analysis, which is subject to the aforementioned limitations. Furthermore, occluding contours is a 2D representation, while we specifically seek a view-consistent, 3D representation. Occluding contours is \textit{view-inconsistent} by construction, which results in the commonly observed flickering artifact when rendering dense views of a 3D shape \cite{hertzman2024newinsights}. 

Our objective is to \textit{abstract visual shape features into a sparse set of 3D strokes}. We optimize the parameters of a set of B\'ezier curves as our stroke representation, where the number of curves determines the level of abstraction.

We show a motivating example in \cref{fig:cylinder-ex}, where we abstract a cylinder with a partial texture, and compare against a naive solution of back-projecting 2D occluding contour curves.
The naive solution calculates a geometric contour at each view (inset), backprojects the result to 3D curves, and aggregates across all views. The naive solution only accounts for the cylinder geometry in a view-dependent manner, resulting in a dense body of strokes. Importantly, the naive solution cannot handle the texture at all. 

Our solution represents the cylinder volume with sparse vertical curves. Different views of the shape are visually distinguished, yet the cylinder's overall geometry is consistent and clear from any given view. Our method also preserves the high-level pattern of the texture.

To achieve our abstraction, we leverage spatially-conditioned semantic supervision by utilizing the intermediate activations of a 2D pre-trained foundation model (CLIP~\cite{radford2021learning}). Our choice of model is justified by previous work that demonstrates CLIP's strong performance in tasks requiring high-level abstract understanding across different modalities \cite{ha_semantic_2022, vinker_clipascene_2023, hernandez-camara_measuring_2024}. We observe that both different layers and architectures enable different levels of control over geometry and texture elements, so we split our optimization into separate geometry and texture abstraction stages.

Our spatial conditioning comes from 3D keypoints, which determine the weight specific visual features are given in our abstraction. The keypoints being in 3D ensures their multi-view consistency and enables fine-grained user control. We show in \cref{fig:keypoint_control} an application where users can iteratively add detail to the abstraction through selecting keypoints corresponding to salient features.

We encourage curve adherence to the input geometry using a neural SDF loss. The surface compliance enables an application where the curves are used as intuitive deformation handles, shown in \cref{fig:deformation} and the supplemental. Our curves' effectiveness as control handles stems from their alignment to semantically meaningful regions of the surface and texture.

In summary, we present a novel technique for abstracting 3D shapes through a sparse set of visually-informed 3D curves. \ourmethod{} can abstract a myriad of shapes from different domains with various visual concepts, geometric structures, and textures. The abstractions are sparse yet effective and maintain high fidelity across arbitrary views. \richard{Our novel localized weighting framework enables interactive user control over the features represented in the abstraction, and our adherence to the input geometry allows for the curves' use as intuitive deformation handles. We encourage the reader to examine the supplemental material, which contains 360-degree videos of our results and an interactive demo of the deformation application using our optimized curves. We plan to release the code for the method and proposed applications in the near future.}


%% file: sec/2_related_work.tex
\begin{figure*}
    \centering
    \includegraphics[width=\linewidth]{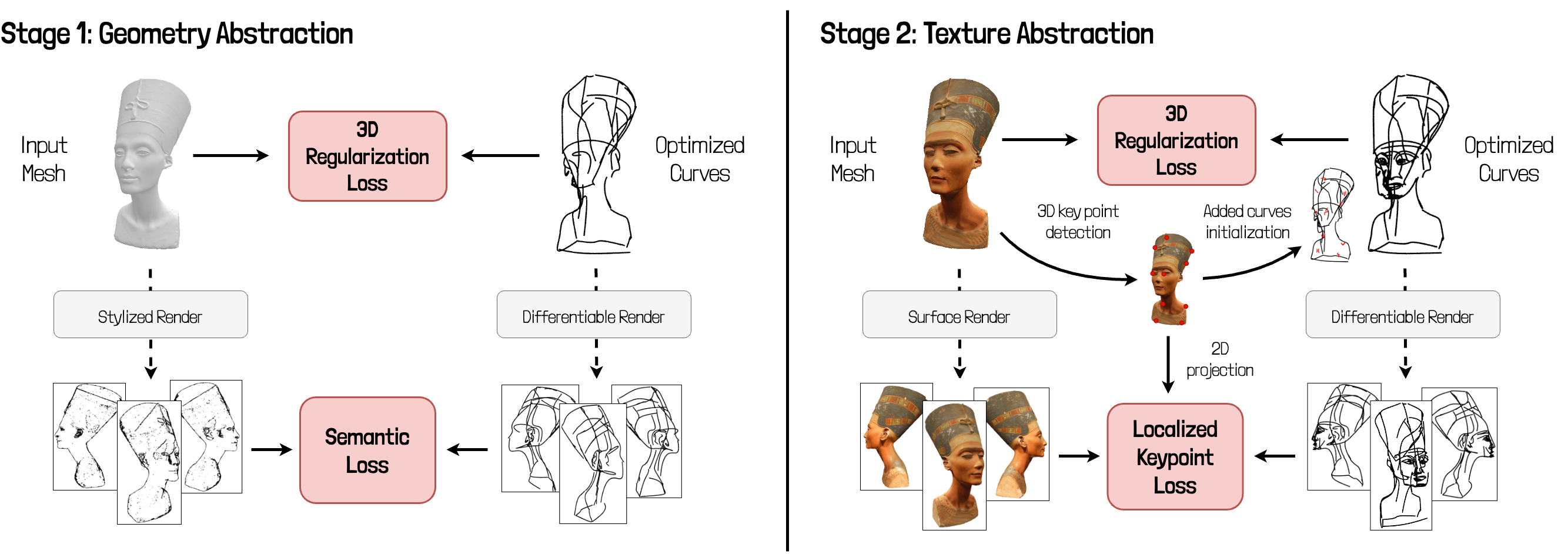}
    \caption{\textbf{Method overview.} In the first stage, our \ourmethod{} learns to abstract the underlying geometry of the shape. In the second stage, we freeze the curves from \il{the first} stage and add new curves that are optimized to represent the shape's texture.}
    \label{fig:pipeline}
\end{figure*}

\section{Related Work}
\label{sec:relatedwork}

\noindent \emph{Shape decomposition.} Shape decomposition is a long-standing problem in 3D geometry analysis, where the goal is to represent the shape by a set of elements, such as primitives \cite{tulsiani2017learning, wu2022primitive, liu2023marching, thiery_sphere-meshes_2013}, convex parts \cite{muntoni2018axis, deng2020cvxnet, jones2022shred, pearl2022geocode}, or Gaussians \cite{hertz2020pointgmm, hertz2022spaghetti}. Tulsiani \etal \cite{tulsiani2017learning} assemble 3D objects from cuboids and obtained simple shape abstractions with consistent structure. In Cvxnet \cite{deng2020cvxnet}, the authors reconstruct a 3D shape by a collection of convex polytopes. Paschalidou \etal \cite{paschalidou2021neural} extend this approach by learning the decomposition and the parts for a set of domain-specific meshes.

This line of work aims at \textit{reconstruction}, whereas our goal is \textit{abstraction}. Moreover, our method does not depend on a dataset and is not limited to a specific domain. It is robust to shapes of arbitrary quality and complexity.
\vspace{-1em}

\paragraph{Non-photorealistic rendering.} A classic problem in graphics is identifying visual contours from 3D geometry to create non-photorealistic renderings. A popular version of this problem is occluding contours \cite{marr1977analysis, koenderink1984does, seales1992viewpoint, boyer19973d, decarlo_suggestive_2003, benard_line_2019}, in which contours representing visible shape regions are delineated from occluded ones. In the classic setting, visible contours are exactly defined to mean the visible surface points tangent to the view vector. Recent efforts have aimed at improving view-consistency and alignment with professional artists \cite{capouellez2023algebraic}, and some apply neural techniques \cite{liu2020neural, liu2021neural}. While such works can successfully depict visual features based on the shape geometry, they lack 3D consistency, and frequently suffer from flickering artifacts \citep{hertzman2024newinsights}.

A recent paper \cite{ye2023nef} and its follow-up \cite{li_3d_2024} introduce implicit edge fields for 3D curve reconstruction from posed images. These works emphasize geometric boundaries, which is distinct from our focus on \emph{abstraction}. A high-level abstraction may reduce a shape feature to a single stroke (e.g. \cref{fig:qualitative-comparison} chair), which is not achievable by these methods. Furthermore, these methods cannot account for texture features which do not induce a prominent edge map. 

\noindent\paragraph{Curve-based abstraction.} Our work is in the domain of sketch abstraction, which aims to depict a scene through 2D or 3D curve primitives. \cite{mehra2009abstraction, hsiao2018multi,gal2009iwires, vinker2022clipasso, choi20243doodle}. Most prior work is based on 2D sketch abstractions \citep{chen2001sketch, photosketch2019, qi_sketchlattice_2021, frans_clipdraw_2022, vinker2022clipasso, tian_modern_2022, song_learning_2018, hahnlein_cad2sketch_2022, para_sketchgen_2021, gal_breathing_2024, vinker_clipascene_2023}, though several works have aimed to represent 2D images through a 3D structures by projection along orthogonal axes \cite{mitra2009shadow, hsiao2018multi, qu2024wired}. Our task is different in that we aim to represent a single shape from all possible viewing directions, not simply three orthogonal views. 

Another line of work aims to reconstruct fabricable 3D wire structures from different modalities, including images \citep{Liu:WireArt:2017, liu_curvefusion_2018, choi_online_2023}, video \citep{wang_vid2curve_2020}, and 3D data \citep{cao_wireframenet_2023, tojo_fabricable_2024, yang_wireroom_2021}. These works are constrained by the fabrication objective, and thus do not abstract fine-grained visual details. 

A final strain of literature deals with analysis of 2D sketches of 3D structures to identify part information and aid in user sketch generation \citep{slzxgm_conceptSketch_sigg13, hennessey_how2sketch_2017, pandey_juxtaform_2023}.  

A recent work proposed optimizing strokes in 3D to match a text description or a guidance image \cite{zhang2024diff3ds}. This work is generative in nature (\eg text-to-3D sketch) and does not involve spatial nor 3D inputs as in our work. \richard{3Doodle \cite{choi20243doodle} presents a technique for optimizing view-dependent and view-independent curves to represent a set of multi-view images. Our work is orthogonal in spirit -- we aim to use a set of view-independent curves for 3D \emph{abstraction} of an \textit{input shape}, and we build our framework leveraging the geometric and semantic shape information. Notably, our use of spatially-driven guidance and SDF loss enables our detail control and deformation applications.}

We use a variant of 3Doodle with only view-independent curves as our main baseline and show results in \cref{fig:qualitative-comparison}. 

%% file: sec/3_method.tex
\section{Method}

\ourmethod{} optimizes a set of 3D cubic B\'ezier curves to abstract a target (potentially textured) shape from all viewing angles. The method takes as input a 3D model and optional user-selected keypoints on the surface. When no keypoints are provided, we automatically detect keypoints using latent backprojection and clustering (\cref{sec:twostage}). 

\subsection{Curve Representation}
\label{sec:curve-rep}
Our 3D strokes are modeled as a set of cubic B\'ezier curves $\{B_i\}_{i=1}^n$, with control points $B_i = (p_i^0, p_i^1, p_i^2, p_i^3) \text{, } p_i^j \in \mathbb{R}^3.$
Points on the curve are sampled through polynomial interpolation of the four control points
\begin{align}\label{eq:bezier}
    B(t) = &(1-t)^3p^0 + (1-t)^2t p^1 \\
    &+ (1-t)t^2 p^2 + t^3p^3 \nonumber
\end{align}
where $0 \leq t \leq 1$.

We make the same assumption as 3Doodle \citep{choi20243doodle} that the camera is sufficiently far from the shape such that orthographic and perspective projection are nearly identical. 
Theorem 1 from 3Doodle thus applies to our method, which establishes equivalence between the normal 3D B\'ezier curves we optimize and the space of 2D rational B\'ezier curves, generated from projecting the 3D curve control points into 2D. Our pipeline renders the 3D B\'ezier curves by first perspective-projecting the 3D control points, then rasterizing the 2D cubic B\'ezier defined from the projected control points (interpolated as in \cref{eq:bezier}). This process is encapsulated by the ``Differentiable Render" step in \cref{fig:pipeline}. The differentiable rasterizer we use is DiffVG \citep{Li:2020:DVG}. 

The resulting image from this rasterization process is $I_{\text{curve}} = \mathcal{R}(\{B_i\})$. We refer to target shape renders as $I_{\text{target}}$.

\subsection{Losses} 
\label{sec:losses}
We leverage the priors of 2D pretrained image encoders to define our semantic losses. Specifically, we design specialized perceptual losses using CLIP \citep{radford2021learning} and LPIPS \citep{zhang2018perceptual} to encourage our rendered 3D strokes to visually match the corresponding renders of the target shape.

The basic structure of our semantic loss is adapted from CLIPasso \citep{vinker2022clipasso}, which compares both the intermediate spatial activations and final global activations between the rendered strokes and the target shape: 
\begin{align}\label{eq:semantic}
    \mathcal{L}_{\text{semantic}} = &\lambda_{\text{fc}} \text{dist}(\text{CLIP}(I_{\text{curve}}), \text{CLIP}(I_{\text{target}})) \\ &+ \sum_{l=3,4} || \text{CLIP}_l(I_{\text{curve}}) - \text{CLIP}_l(I_{\text{target}})||_2^2 \nonumber
\end{align}
where $\text{dist}(\cdot,\cdot)$ measures cosine similarity, $\text{CLIP}$ is the global CLIP encoding, and $\text{CLIP}_{l}$ is the layer $l$ activation. 

We find this semantic loss, however, to be lacking when it comes to representing fine details, including textures. Thus, we introduce a \textit{spatial weighting} framework, which directs the optimization towards specific visual features. 
\begin{figure}
    \centering
\includegraphics[width=\linewidth]{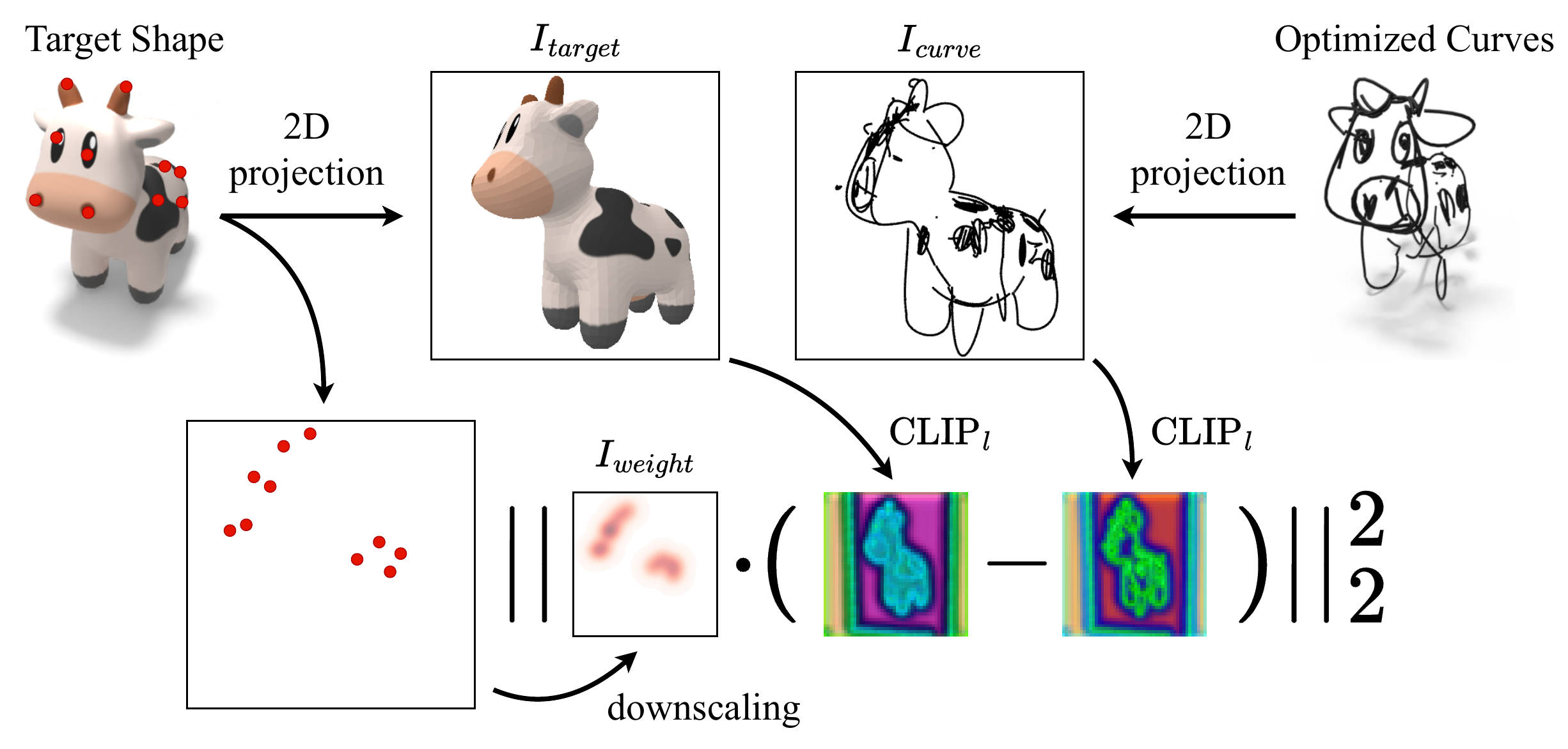}
    \caption{\textbf{Localized keypoint loss.} Our localized keypoints weight the loss between the intermediate feature maps of the encoded curve render $I_{\text{curve}}$ and the target shape render $I_{\text{target}}$. This weight is obtained through projecting 3D keypoints (red), followed by a Gaussian filter to obtain the weight map $I_{\text{weight}}$. This loss focuses the optimization on visual features local to the keypoint.}
    \label{fig:spatial}
    \vspace{-1em}
\end{figure}
\paragraph{Localized Keypoint Loss.} Our spatial weighting is based on previous work that establishes the spatial correlation between CLIP's intermediate activations and the input image. \citep{gilurispatialcorrelation2019}. Features of interest can be emphasized in these intermediate layers by identifying their location in the render and tracing the correspondence to the downsampled feature maps, hence \emph{localizing} the features in the feature maps.

We assume as optional input user-defined 3D keypoints. If no input keypoints are given then our method automatically detects keypoints, as described in \cref{sec:twostage} and the supplemental. These keypoints should indicate salient visual features on the input shape or texture, and we leverage that information to guide abstraction of the specific features. 

For every sampled view during optimization, we project the keypoints to the same views and identify their positions in the corresponding renders. Once we have the keypoint locations in the rendered image, we construct a weight map at the same resolution as the image based on a Gaussian dropoff from the keypoint center. Specifically, we construct the weight image $I_{\text{weight}}$ such that

\begin{equation}\label{eq:semanticweight}
    I_{\text{weight}}(x, y) = 1 + \sum_{p} e^{\frac{-||(x,y) - p||^2}{2\sigma^2}}
\end{equation}
where $x, y \in [0,1]$ index the normalized image pixels, and $p \in [0,1]$ is the keypoint projected to normalized coordinates. We add 1 to the weights so that regions far from the keypoints still contribute to the loss. We use these constructed weight maps to weight our semantic loss \cref{eq:semantic} for each render, such that each L2 term in the intermediate layer losses $||\text{CLIP}_l(I_{\text{curve}}) - \text{CLIP}_l(I_{\text{target}})||_2^2$ becomes $||I_{\text{weight}} \cdot (\text{CLIP}_l(I_{\text{curve}}) - \text{CLIP}_l(I_{\text{target}}))||_2^2$. The weight map is downsampled to match the resolution of each intermediate feature map. We visualize this process in \cref{fig:spatial}.

We maintain a z-buffer for the shape renders, such that if a keypoint's projected depth is higher than the shape depth, then the keypoint is occluded by the surface from that view and does not contribute to the weight image. This prevents keypoints from being attributed to incorrect surface regions.

To provide additional structural constraints, we include an LPIPS loss term \citep{zhang2018perceptual}. LPIPS is a popular perceptual loss function which is known to be sensitive to geometric layouts \citep{choi20243doodle}. The final localized keypoint loss becomes:
\begin{align}\label{eq:spatial}
    \mathcal{L}_{\text{local}} & = \lambda_{\text{fc}}\bar{I}_{\text{weight}} \text{dist}(\text{CLIP}(I_{\text{curve}}), \text{CLIP}(I_{\text{target}})) \\ &+ \sum_{l=3,4}|| I_{\text{weight}} \cdot (\text{CLIP}_l(I_{\text{curve}}) - \text{CLIP}_l(I_{\text{target}}))||_2^2 \nonumber \\ &+ \lambda_{\text{lpips}}\text{LPIPS}(I_{\text{curve}}, I_{\text{target}}) \nonumber
\end{align}
where $\bar{I}_{\text{weight}}$ indicates the mean-pooled weights.

\paragraph{SDF Regularization.} To encourage adherence of the curves to the target shape geometry, we use a loss based on the shape's Signed Distance Field (SDF). We fit an MLP $\Phi$ on the shape's SDF (Unsigned Distance Field in the case of shapes with boundaries) to obtain a neural SDF. During stroke optimization, we densely sample each 3D curve and query their SDF values using the neural SDF, penalizing values outside of the zero level set:
\begin{equation}\label{eq:sdf}
    \mathcal{L}_{\text{SDF}} = \frac{1}{n \cdot k} \sum_{i=1}^n \sum_{k = 1}^s |\phi(B_i(t_k)|
\end{equation}
where $t_k \in [0,1]$ are random samples along the B\'ezier curve for $s$ total samples. This loss helps to anchor abstracted features to the surface implied by the curve set.

\paragraph{View Regularization.} We further regularize the abstraction by enforcing that all curves are visible from all sampled viewing angles, which prevents curves from hiding outside of the camera view. With $\mathcal{P}$ indicating the perspective projection of the B\'ezier curve control keypoints, we have: 
\begin{align}\label{eq:ndc}
    \mathcal{L}_{\text{ndc}} &= \sum_{i=1}^n \sum_{t} \text{ReLU}(\mathcal{P}(B_i(t)) - 1) + \text{ReLU}(-\mathcal{P}(B_i(t)))
\end{align}
where $\text{ReLU}$ is a Rectified Linear Unit. This penalizes all curve points with image coordinates outside of the render frame (0-1). The first term has non-zero value when the coordinates are greater than 1, and the second term has non-zero value when the coordinates are less than 0.

\subsection{Two-Stage Optimization} 
\label{sec:twostage}
In our experiments we find that different CLIP architectures are sensitive to geometry or semantic shape features. To exploit this, we use a two-stage training pipeline, in which the first stage optimizes for the shape \textit{geometry} and the second stage abstracts the shape \textit{texture}, and each stage leverages different CLIP architectures and semantic losses. 

\paragraph{Keypoint Initialization.} When keypoints are not included in the input, we automatically identify keypoints of interest on the shape's surface using the 2D to 3D feature back-projection method introduced in Backto3D \citep{wimmer2024back} with KMeans clustering \citep{lloyd1982least}. See supplemental for more details. 

\paragraph{Geometry Abstraction.} During stage I optimization, the B\'ezier curves are initialized using furthest point sampling, with the control points drawn from small Gaussians around each sampled point. These curves are optimized towards the shape geometry with a combination of our original (non-localized) semantic (\ref{eq:semantic}), SDF (\ref{eq:sdf}), and NDC (\ref{eq:ndc}) losses, supervised with Freestyle renders (see supplemental) $I_{\text{target}}^{\text{free}}$ of the target shape. The stage I loss is:
\begin{align*}
    \mathcal{L}_{\text{I }} = &\mathcal{L}_{\text{semantic}}(I_{\text{curve}}, I_{\text{target}}^{\text{free}}) + 0.1 \cdot \mathcal{L}_{\text{SDF }} + \mathcal{L}_{\text{ndc}}(\{B_i\}).
\end{align*}
\paragraph{Texture Abstraction.} In our second stage, we freeze the set of geometry curves optimized in the first stage and initialize a new set of curves in the same way as the first stage, using furthest distance sampling. These curves are then optimized to represent the semantic texture of the shape using our localized keypoint loss (\ref{eq:spatial}), SDF (\ref{eq:sdf}), and NDC (\ref{eq:ndc}) losses, supervised with surface (potentially textured) renders of the shape $I_{\text{target}}^{\text{surface}}$. The stage II loss is: 
\begin{align*}
    \mathcal{L}_{\text{II }} = &\mathcal{L}_{\text{local}}(I_{\text{curve}}, I_{\text{target}}^{\text{surface}}) + \mathcal{L}_{\text{SDF}} + \mathcal{L}_{\text{ndc}}(\{B_i\}).
\end{align*}
We use CLIP architectures RN101 and RN50x64 for $\mathcal{L}_{\text{semantic}}$ and $\mathcal{L}_{\text{local}}$, respectively. We've found empirically that RN101 tends to be more sensitive to geometric structures, whereas RN50x64 is more sensitive to higher-level visual concepts.

%% file: sec/4_experiments.tex
\section{Experiments}
\label{sec:experiments}


We evaluate \ourmethod{} across a wide variety of shapes and demonstrate multi-view fidelity and abstraction control in \cref{sec:qualitative}. We compare \ourmethod{} quantitatively and qualitatively to relevant baselines in \cref{sec:baselines}. Finally, \cref{sec:applications} showcases the interactive feature control and shape deformation applications enabled by our method.

The shapes in our experiments are aggregated from COSEG \cite{kaick2011prior}, the Meta Digital Twin Catalog \citep{pan2023aria}, and Keenan Crane's 3D model repository \citep{crane2013robust}. \ourmethod{} is robust to meshes with varying topology, complexity and quality.

For all experiments, we randomly sample views spanning 0 to 30 degrees elevation and 0 to 360 degrees azimuth. We use 20 curves for stage 1 and initialize 20 additional curves for stage 2. 

\subsection{Qualitative Results} \label{sec:qualitative}

\begin{figure}
    \centering
    \includegraphics[width=\linewidth, trim=0 0 0 0, clip]{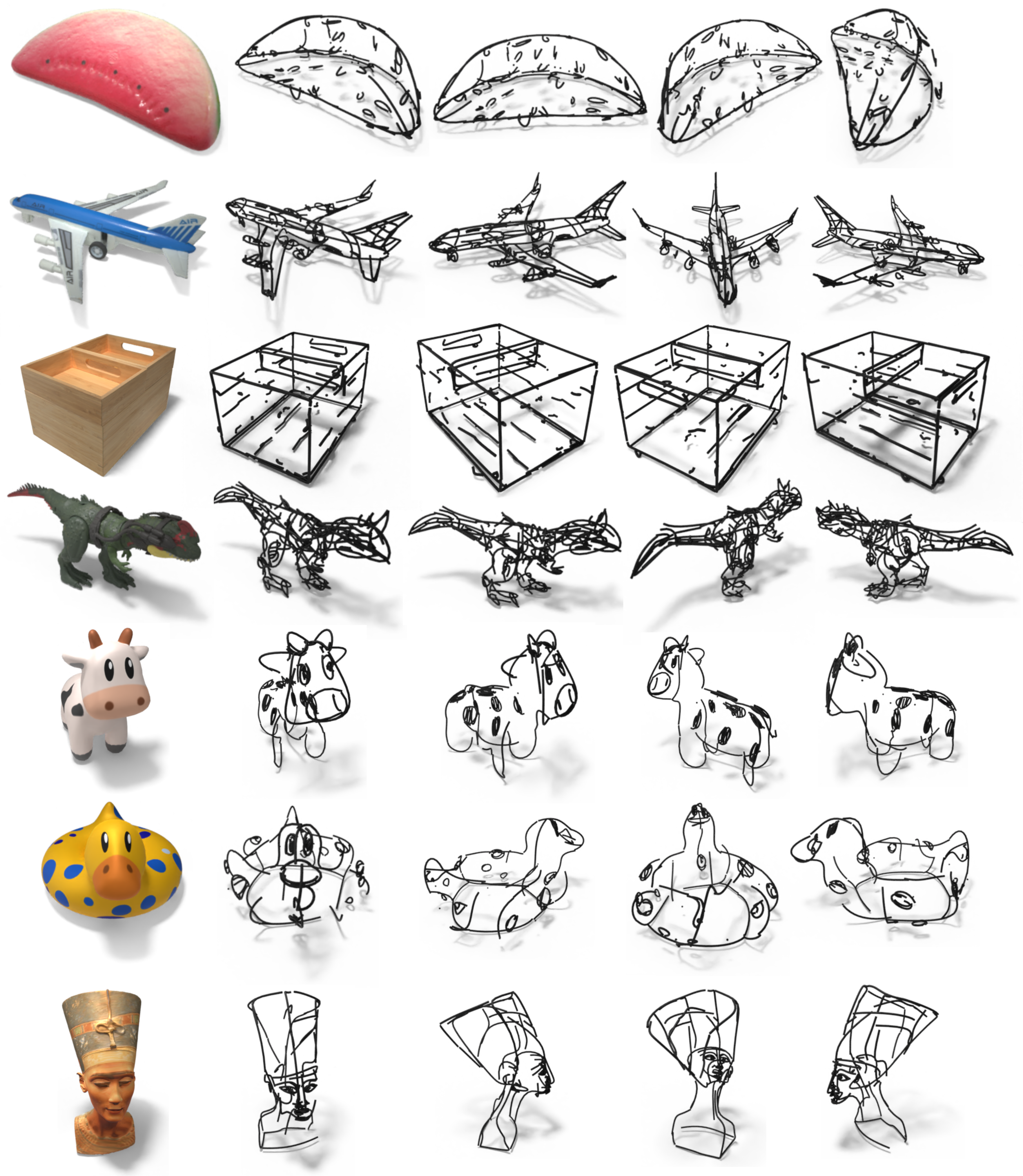}
    \caption{\textbf{Qualitative results for textured objects.} We show \ourmethod's the result on a collection of textured meshes.}
    \label{fig:textured}
\end{figure}

\begin{figure}
    \centering
    \includegraphics[width=\linewidth, trim=0 0 0 0, clip]{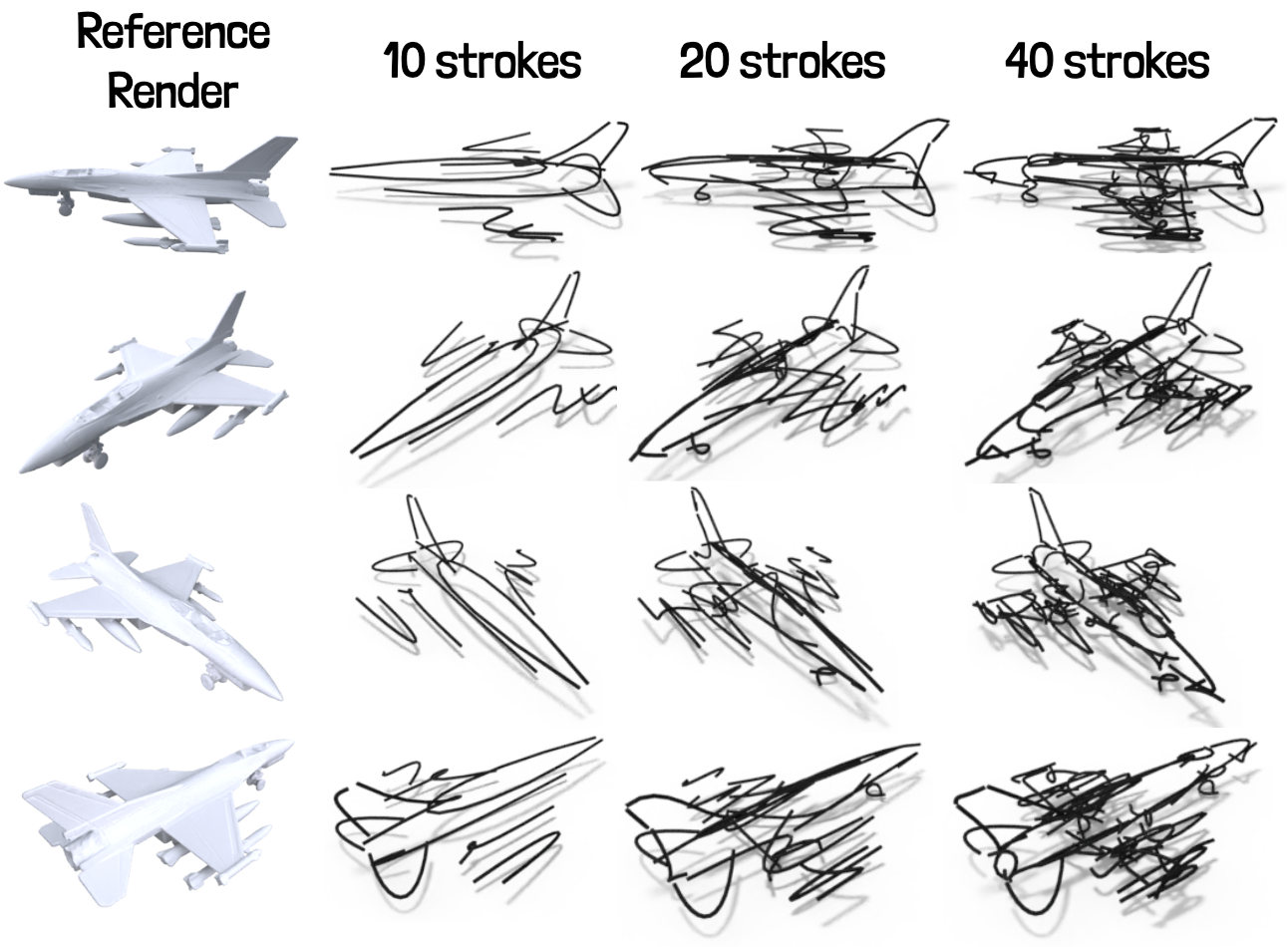}
    \caption{\textbf{Abstraction control.} The level of abstraction is implicitly controlled by the number of curves. As the curve count grows, the abstraction captures more fine details.}
    \label{fig:ncurves}
\end{figure}

\paragraph{Shape Abstractions.} \cref{fig:textured} shows 3D stroke abstractions generated by \ourmethod{} for textured shapes.\ourmethod{} successfully abstracts textures and small salient structures, such as watermelon seeds (Row 1) and spots (Rows 5/6), as well as facial features (Rows 5/7). We show results on untextured shapes in \cref{fig:untextured} in the supplemental, which demonstrate that our method can also represent complex geometries.
\vspace{-1em}

\paragraph{Abstraction Control.} Our method automatically adapts the level of abstraction based on the number of strokes being optimized. \cref{fig:ncurves} shows that adding more strokes adds progressvely more detail to the abstraction, and our method produces high quality abstractions across all levels. 
\vspace{-1em}

\begin{figure}[h]
    \centering    \includegraphics[width=\linewidth]{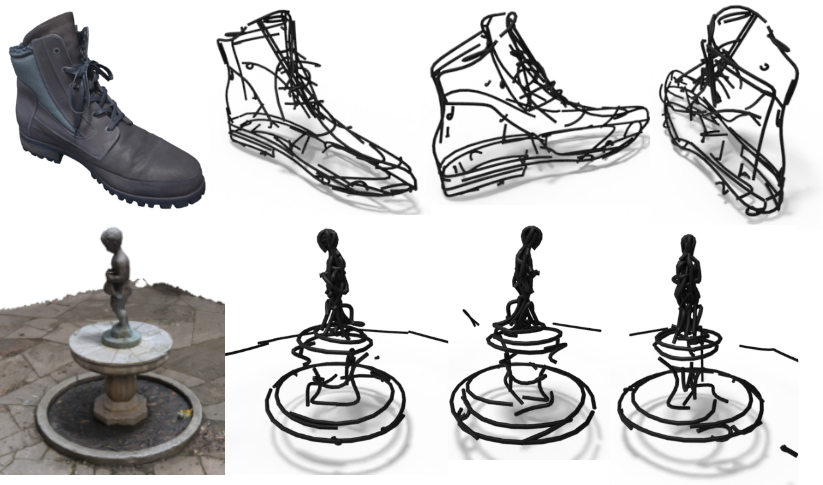}
    \caption{\textbf{In-The-Wild Reconstructions.} We abstract two shapes from the Real-World-Textured Things dataset \cite{maggiordomo_real-world_2020}, which provides textured models reconstructed using modern off-the-shelf photo-reconstruction tools on in-the-wild images. The shoe (top) contains 360 boundary edges, 31 non-manifold edges, and 9 non-manifold vertices. The sculpture (bottom) contains 113601 boundary edges and 1274 non-manifold edges.}
    \label{fig:inthewild}
\end{figure}

\paragraph{In-The-Wild Abstraction.} \ourmethod{} has no requirements on the quality of the input geometry. Only the Freestyle rendering will be affected by poor geometry, but CLIP supervision is robust to noisy render artifacts due to its global semantic understanding.  We show in \cref{fig:inthewild} that WIR3D successfully abstracts shapes built from photo reconstruction software, which have many geometric defects. 

\subsection{Baseline Comparisons} \label{sec:baselines}

The baselines we compare to are NEF \cite{ye2023nef} and 3Doodle \cite{choi20243doodle}. For both methods, we use the publicly available repositories published by the respective authors. We standardize the sampled views and use the default hyperparameters otherwise. 3Doodle initializes its curves using SFM, but given the method's sensitivity to poor initialization, we standardize initialization to the same furthest point samples as our method, finding that they perform better than SFM. We use 40 curves for 3Doodle to match \ourmethod{}, but NEF has no ability to determine the curve count. 

NEF often struggles to fit reasonable point clouds, which results in meaningless curves. We show qualitative comparison to NEF in the supplementary, where we train NEF on a set of viewpoints optimized for the method. 
\vspace{-1em}

\begin{figure}
    \centering
    \includegraphics[width=\linewidth, trim=0 0 0 0, clip]{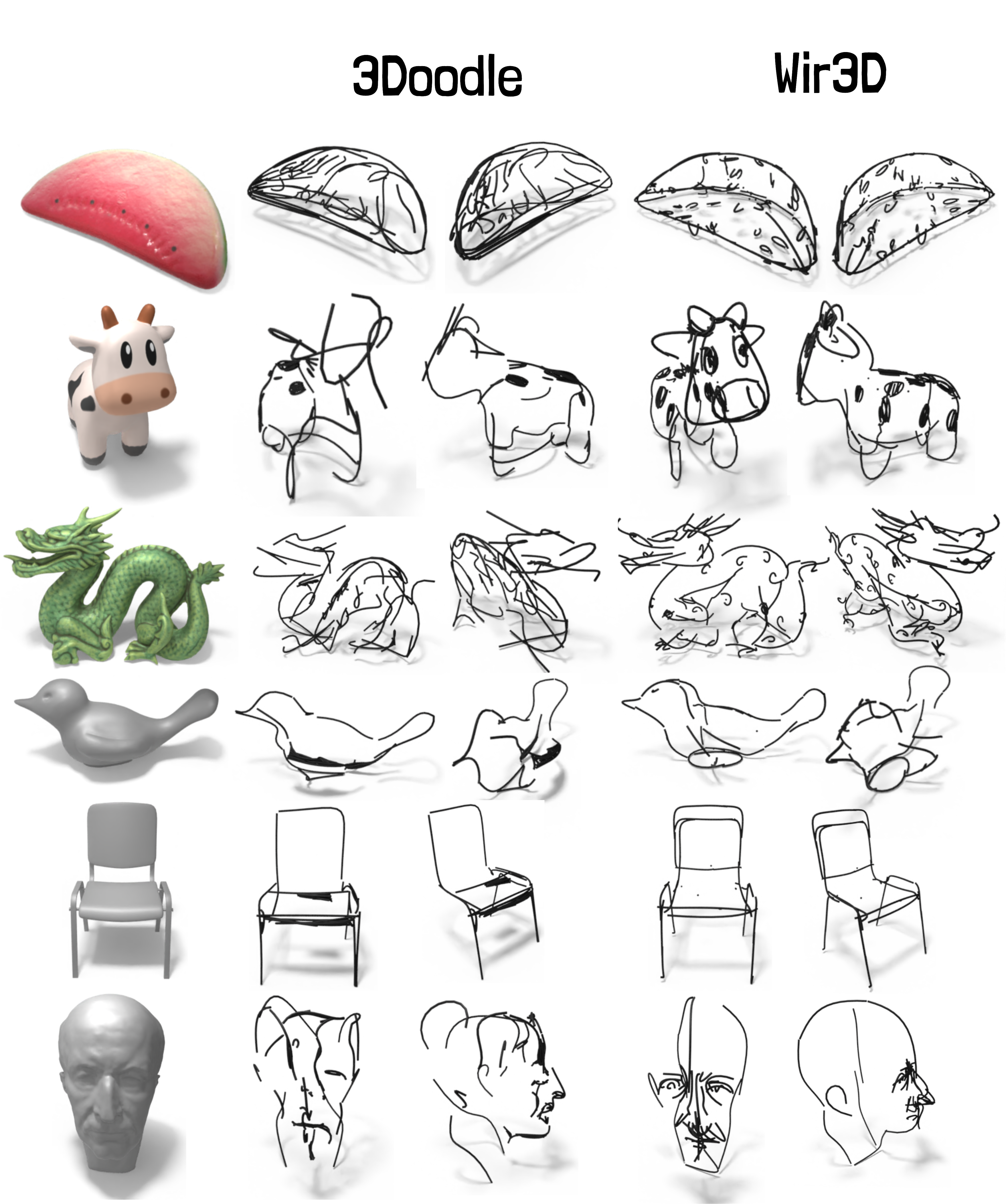}
    \caption{\textbf{Qualitative comparison.} We compare a subset of our textured and untextured results against 3Doodle. 3Doodle produces reasonable abstractions that capture the overall structure but lack precision when it comes to the feature details.}
    \label{fig:qualitative-comparison}
\end{figure}

\paragraph{Qualitative Comparison.} \cref{fig:qualitative-comparison} compares against \ourmethod{} against 3Doodle for both textured and untextured shapes. Our method consistently captures the global geometry, whereas 3Doodle struggles in a low signal context such as untextured shapes (missing back left chair leg, flattened bird geometry). 3Doodle is consistently insensitive to facial structures, whereas our localized weighting ensures that we abstract them (rows 2, 6). Similarly, we are able to represent fine-grained texture structures as high-level visual motifs, such as the dragon scales (row 3).
\vspace{-1em}

\begin{table}
    \centering
    \footnotesize
    \addtolength{\tabcolsep}{-0.4em}
    \begin{tabular}{lcccc}
    \toprule
    Method & LPIPS ($\downarrow$) & $\text{CLIP}^{\text{img}}$ ($\uparrow$) & User Rank ($\uparrow$) & Coverage ($\downarrow$)\\
    \midrule
    NEF & 0.313 & 0.86 & - & 0.056 \\ 
    3Doodle & 0.246  & 0.900 & 0.12 & 0.020 \\
    \ourmethod{} (Ours) & \textbf{0.227} & \textbf{0.909} & \textbf{0.88} & \textbf{0.008}\\
    \bottomrule
    \end{tabular}
    \caption{We compare \ourmethod{}, 3Doodle \cite{choi20243doodle}, and NEF \cite{ye2023nef} through 3 perceptual metrics computed over novel views of the curves and the target shape. ``LPIPS" measures the average LPIPS similarity score using AlexNet. ``$\text{CLIP}^{\text{img}}$" measures the average cosine similarity scaled between (0,1) using the ViT/B-32 model. ``User Rank" measures the percentage of user responses that prefer one method over the other ($N=96$). We also compute a geometric metric ``Coverage", which measures the 1-direction Chamfer distance between the curves and the target 3D surface.}
    \label{tab:quantitative}
\end{table}

\paragraph{Quantitative Comparison.} We use three perceptual metrics and one geometric metric, reported in \cref{tab:quantitative}. The first two, LPIPS \citep{zhang2018perceptual} and $\text{CLIP}^{\text{img}}$ \citep{wang2022exploring}, are the same metrics reported by 3Doodle. Our method outperforms 3Doodle on both metrics, though numerically our $\text{CLIP}^{\text{img}}$ is not much higher. We note that CLIP has been shown to be insensitive to large visual differences, as systematized in \cite{tong_eyes_2024, wang_diffusion_2024}. 
We further illustrate this point in \cref{fig:metricrobustness} (supplemental), where 
we show that a randomly oriented cow outline obtains a similar $\text{CLIP}^{\text{img}}$ score as the edge map \citep{canny_computational_1986} of the \textit{ground-truth render}. Thus, the numerical difference in $\text{CLIP}^{\text{img}}$ is an unreliable indicator of the difference in visual quality. 

This motivates our user perceptual study ($N=96$), which takes all the results from our dataset and compares Wir3D results against 3Doodle. Specifically, we display rotating gifs of both results side by side, along with a rotating gif of the target shape, and ask users to rank the 3D strokes based on how well they represent the target shape. We collected responses from 96 users and compute the frequency our method is ranked higher than 3Doodle. Our curves are ranked higher \textbf{88\%} of the time. Screenshots from the study are shown in the supplemental.

Our final metric, ``Coverage", is a geometric metric that quantifies how well the stroke abstractions cover the original surface. We measure coverage by sampling 100k surface points  and computing the 1-direction Chamfer distance from the points to the strokes. This evaluates whether each point on the surface has a curve reasonably close to it. Our method significantly outperforms the baselines in terms of coverage, with a $>$2x reduction in coverage distance relative to 3Doodle, and $>$4x reduction relative to NEF. 

\subsection{Ablations} \label{sec:ablations}

\begin{table}[t]
    \centering
    \small
    \begin{tabular}{lcccc}
    \toprule
    Method & LPIPS ($\downarrow$) & $\text{CLIP}^{\text{img}}$ ($\uparrow$) & Coverage ($\downarrow$) \\
    \midrule
    \ourmethod{} & \textbf{0.227} & \textbf{0.909} & \textbf{0.008} \\
    No SDF & 0.229 & 0.904 & 0.012 \\
    No Local & 0.233 & 0.905 & 0.009 \\
    w/o Stage 1 & 0.248 & 0.900 & 0.016 \\
    No CLIP Layers & 0.294 & 0.891 & 0.012 \\
    \bottomrule
    \end{tabular}
    \caption{\textbf{Ablation Quantitative Metrics.} ``WIR3D" is our full method. ``w/o Stage 1" is the Stage 1 ablation. ``No CLIP layers" is the ablation on CLIP intermediate layer activations. ``No SDF" is the ablation on the SDF loss. ``No Local" is the ablation on the localized keypoint loss.}
    \label{tab:ablquantitative}
\end{table}

We perform a thorough ablation study demonstrating the importance of our design choices in \ourmethod{}. Figures for each ablation are shown in the supplemental. We report the ablation metrics in \cref{tab:ablquantitative}. Removing the CLIP intermediate layers (``No CLIP Layers'') from our supervision results in the highest reduction in quality (supplemental \cref{fig:ablfc}), followed by removing the stage 1 training (``w/o Stage 1''). Removing this stage results in the optimization biasing towards certain views and creating an overall ``flattened" effect of the geometry (supplemental \cref{fig:ablstage1}). Removing the localized keypoint loss (``No Local'') also results in some reduction in the metrics, but as discussed in \cref{sec:baselines}, perceptual metrics may not be sensitive to the fine visual detail this loss is designed to capture. Removing the SDF loss results in worse geometric coverage, as expected (``No SDF'').  
\vspace{-2em}

\begin{figure}[h]
    \centering
    \includegraphics[width=0.9\linewidth]{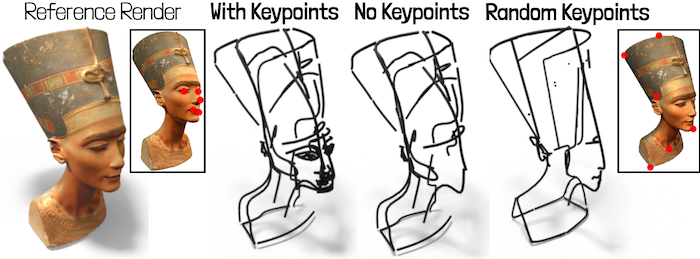}
    \caption{\textbf{Keypoint Robustness.} Our localized keypoint loss is robust. When keypoints are random (``Random Keypoints''; positions in insets shown), the result is similar to when the localized keypoint loss is not used at all (``No Keypoints'').}
    \label{fig:kpablation}
\end{figure}

\paragraph{Keypoint Robustness.} We test the robustness of our keypoint loss to noisy keypoints. In \cref{fig:kpablation}, we randomly sample keypoints and compare against the original results (``With Keypoints'') and the results without using the localization weighting (``No Keypoints''). When noisy keypoints are given, the results are similar to the results without localization weighting (becomes \cref{eq:semantic}). With meaningful keypoints, our method produces strictly improved results.

\begin{figure*}[t!]
    \centering
    \includegraphics[width=\linewidth]{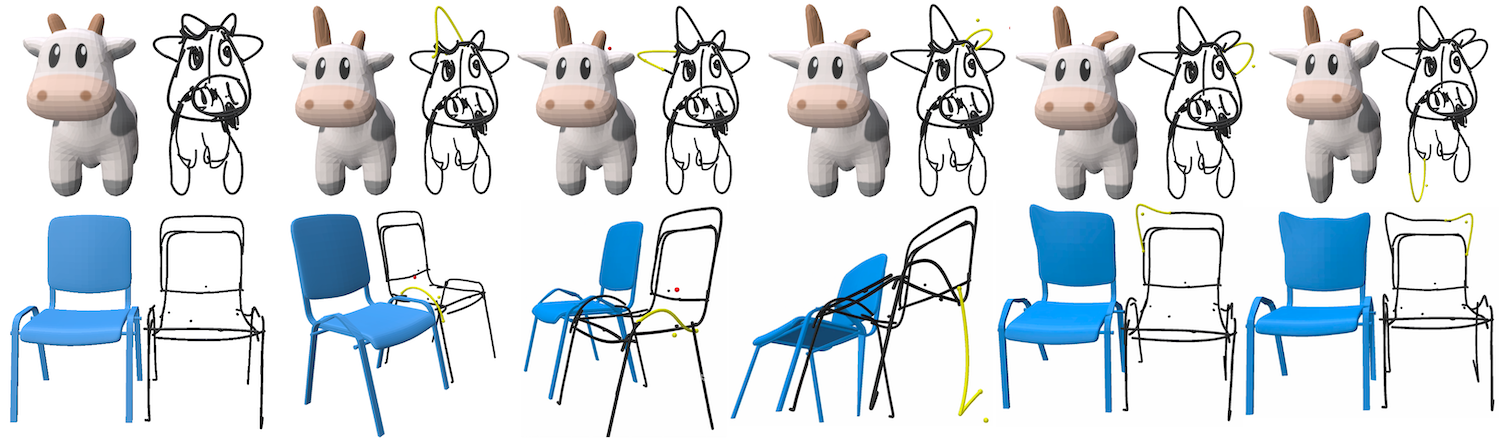}
    \caption{\textbf{Deformation application.} The curve abstractions make for intuitive control handles for shape deformation. The curves wrap salient features such that deformations are transferred smoothly through a simple L2 distance skinning procedure.}
    \label{fig:deformation}
\end{figure*}

\begin{figure}[h!]
\centering
\includegraphics[width=\linewidth, trim=0 0 0 0, clip]{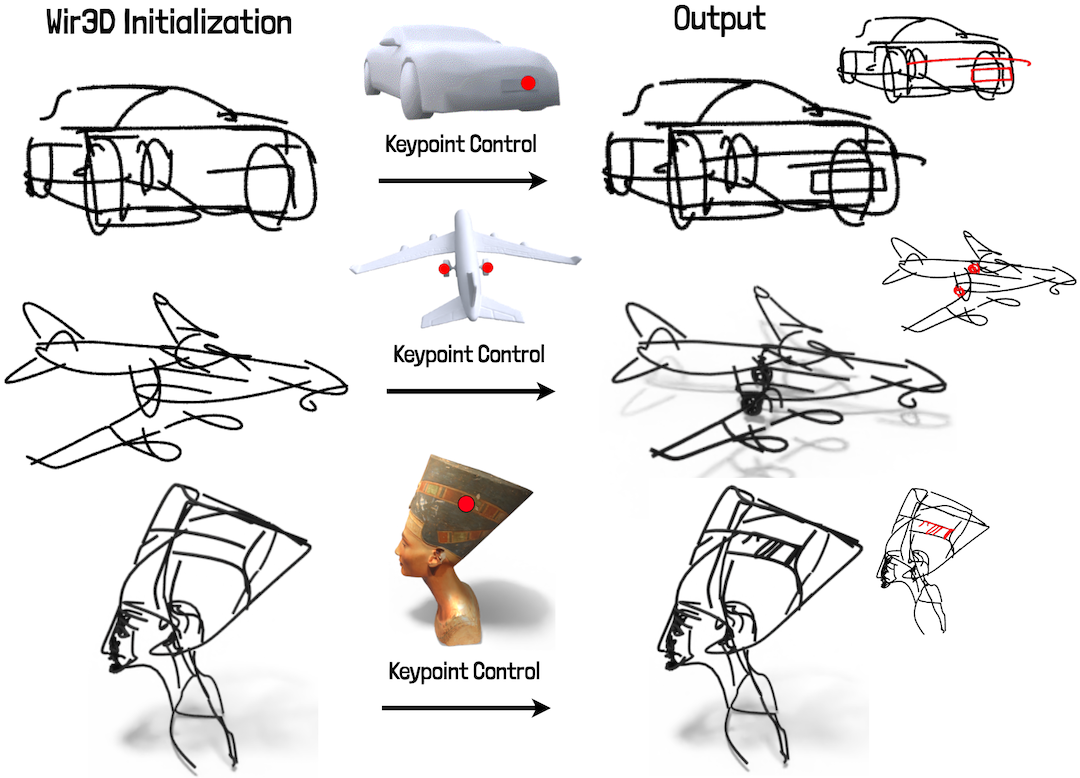}
\caption{\textbf{Keypoint control.} Our spatial weighting framework enables user-interactive detail refinement.}
\label{fig:keypoint_control}
\end{figure}

\subsection{Applications} 
\label{sec:applications}
We demonstrate two applications enabled by our 3D curve representation. The first is user-interactive feature control. By selecting additional keypoints, new curves can be added and quickly optimized to add detail to any feature of interest. We show in \cref{fig:keypoint_control} that keypoints can be used to make structures more explicit (e.g. wheels on the plane) or to add texture detail (e.g. nefertiti headband). The refinement is completed in a few hundred iterations (1 min). 

The curves' adherence to the input geometry makes them intuitive deformation handles. \cref{fig:deformation} shows examples of interactive deformations, where Euclidean distance-based skinning weights map deformations from the curves to the shape surface. The curves naturally wrap salient features so that the deformations are smoothly transferred, and relevant parts are easily manipulated. To verify the effectiveness of these curves as deformation handles, we conduct a perceptual study comparing deformations generated by WIR3D against ARAP \cite{sorkine_as-rigid-as-possible_2007}, a classic deformation method. We drag a single control point on a \ourmethod{} curve and compare the result to dragging the nearest vertex with ARAP. We ask respondents to ``pretend to be 3D modelers in charge of sculpting various shapes'', and ask them which deformation result is most desirable from that perspective. \begin{wrapfigure}{r}[10pt]{0.5\linewidth}
\vspace{-1em}
\hspace{-2em}
\centering
\includegraphics[clip=true, trim={10 10 10 10},width=\linewidth]{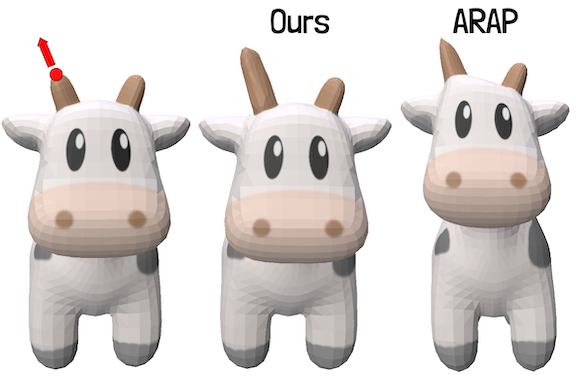}
\vspace{-1em}
\end{wrapfigure}
The respondents (N=42) found our WIR3D-based deformations to be more desirable \textbf{80\%} of the time. We show one example in the inset, with remaining study details in the supplemental. Our curve-based handles preserve the underlying shape semantics better, such as deforming only the cow's horn while leaving its head intact.

%% file: sec/5_conclusion.tex
\section{Conclusion}

\ourmethod{} is a method for 3D model abstraction into sparse strokes, parameterized as cubic B\'ezier curves in 3D. We introduce a novel localized keypoint loss, which allows the abstractions to represent fine-grained geometry and texture details. This localized weighting framework enables user control over the local abstraction detail through interactive keypoint selection and detail refinement. Our method is robust across models of arbitrary topology and   quality. We encourage the 3D curves to maintain close correspondence to the original surface, which can enable intuitive shape editing applications such as curve-based deformation. 

\smallskip
\textit{Limitations.} The quality of \ourmethod{} is conditional on how good the keypoints are (\cref{fig:kpablation}). Another limitation is the preprocessing. Fitting a neural SDF, running our automatic keypoint algorithm, and generating the Freestyle renders may take a lot of time depending on the input (2 hours for Nefertiti at 100k faces).

\section{Acknowledgements}
This research was supported by grant \#2022363 from the United States - Israel Binational Science Foundation (BSF), grant \#2304481 from the National Science Foundation (NSF), and gifts from Adobe, Snap, and Google.

%% file: sec/X_suppl.tex
\clearpage
\setcounter{page}{1}
\maketitlesupplementary
\renewcommand*{\thesection}{\Alph{section}}
\setcounter{section}{0}

\section{Data Preprocessing}
\label{supp:data-preprocess}
We leverage the input surface not just in our SDF regularization, but also in generating supervision data specialized for our task. Specifically, we generate stylized Freestyle renders which isolate the key geometric features of a shape for our stage I optimization. In the case where a user does not supply keypoints, we leverage the priors of 2D foundation models to automatically detect keypoints which correspond to salient shape features. 

\paragraph{Freestyle Rendering} Standard opaque surface renders are not ideal for our curve representation, which are non-occlusive by construction. Optimizing with these surface renders can result in under-detailed abstractions or particular Janusing artifacts \citep{susarla_janus_2023}, where curves positioned on the opposite side of the viewed surface end up being optimized for the wrong side. Furthermore, when the shape is untextured, surface renders may be poor at exhibiting key geometric structures. 

To resolve this, we render the shapes in a stylized fashion to allow for each view to isolate the shape geometric structure and take into account the occluded shape features. Specifically, we render the shapes using the Freestyle rendering engine \citep{decarlo2023suggestive} in Blender and render the shape in terms of its view-dependent contours, \textit{without} accounting for occlusions. These Freestyle renders are purely based on the shape geometry and do not take into account any textures. Thus, these renders are appropriate for the first stage of our optimization (\cref{sec:twostage}) where we focus on capturing the shape geometry. 

\paragraph{Keypoint Detection.} When keypoints are not included in the input, we automatically identify keypoints of interest on the shape's surface using the 3D feature extraction method developed in Backto3D \citep{wimmer2024back}. This method back-projects 2D image features to a 3D shape using a simple averaging scheme. Our assumption, following Backto3D, is that these backprojected features contain meaningful information of the shape's salient visual features, and thus can be leveraged for identifying keypoints relevant to those features. 

Specifically, we render views of the 3D model and encode them using CLIP RN50x16, back-project the pixel-level latent features to shape vertices, and average the features among duplicate vertices captured in different views.

Once we have 3D features on the shape, we apply KMeans clustering over these features \cite{lloyd1982least} and obtain $k$ latent clusters, where $k$ is the number of keypoints we wish to obtain. We interpret these clusters as aggregating surface points with similar visual content. We identify the vertex whose features are closest to the cluster centers as keypoints, since these vertices are most likely to represent the key visual feature associated with the cluster. We make $k$ the number of curves we initialize in stage 2 of our optimization, though this number can be adjusted depending the number of salient elements on the shape. 

\begin{figure}
    \centering
    \includegraphics[width=\linewidth]{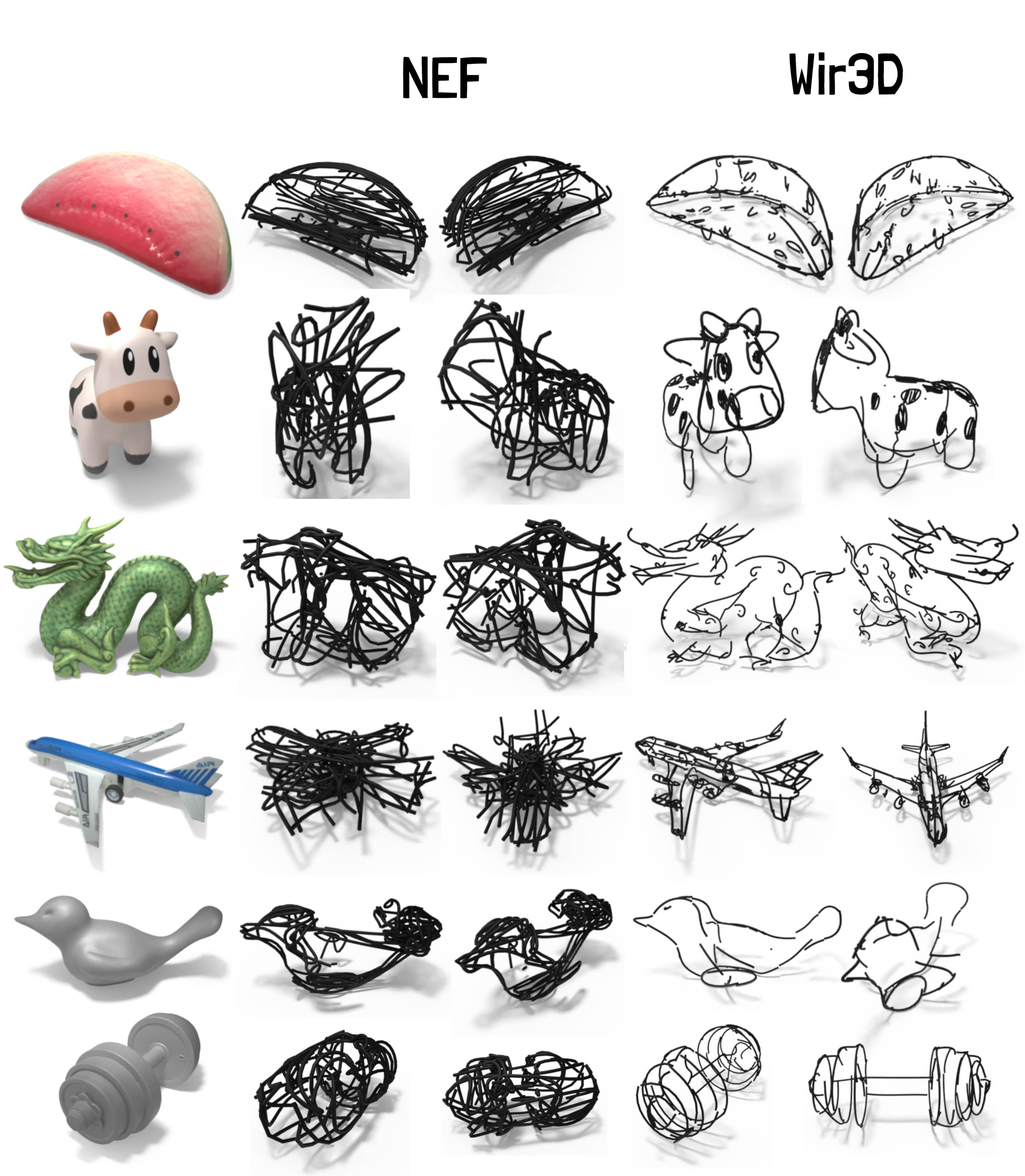}
    \caption{\textbf{NEF qualitative comparison.} We show NEF results on the same models we compare to 3Doodle in the main paper. NEF is specialized for simple manufactured CAD shapes, so it struggles to fit edges to more complex surfaces. This limitation was similarly observed in 3Doodle.}
    \label{supp:nefqualitative}
\end{figure}

\begin{figure}
    \centering
    \includegraphics[width=\linewidth]{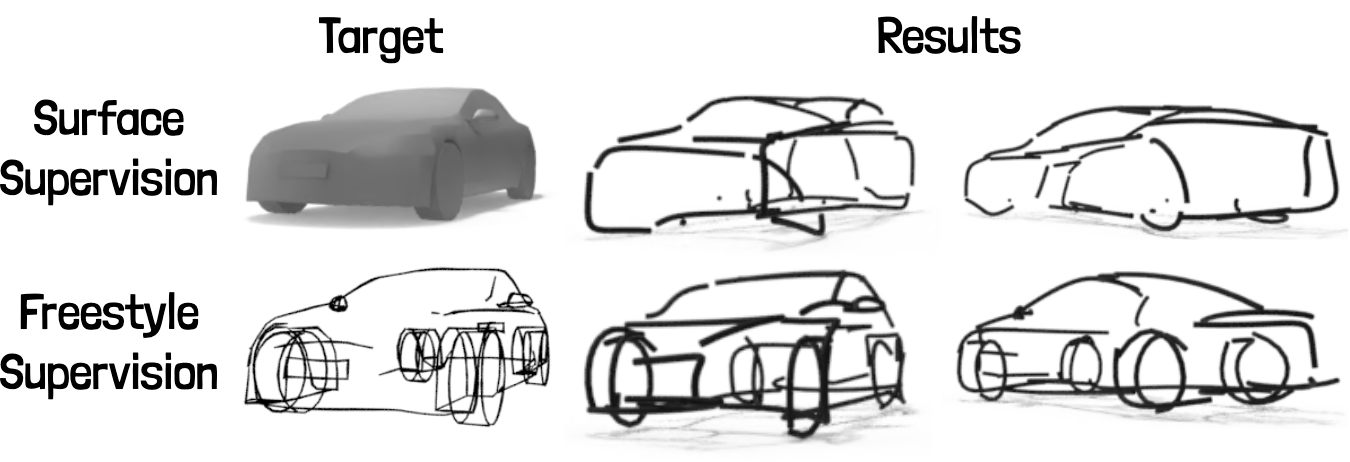}
    \caption{\textbf{Freestyle render ablation.} Running our method without freestyle renders still produces a reasonable abstraction, but key geometric features, such as the wheels of the car, may be missed due to the lack of visual signal from the surface renders.}
    \label{supp:ablfreestyle}
\end{figure}

\section{Neural Edge Field Comparison}
We show a qualitative comparison to NEF in \cref{supp:nefqualitative}, using the same models we show in the main paper for 3Doodle, except for the models which NEF fails to produce meaningful point clouds for. Though NEF can capture the rough silhouette of the target shape, the method is specialized for simple manufactured surfaces with sharp corners, so it struggles to place curves meaningfully on more complex surfaces. This results in a messier and harder-to-identify abstraction.

\section{Perceptual Metric Details.}
The LPIPS metric is based on an AlexNet architecture trained for image classification fine-tuned with a linear layer on an annotated perceptual similarity dataset. Notably, we use the VGG variant of LPIPS for optimization, which is a commonly performed split between optimization and evaluation, and is similarly done in 3Doodle.

$\text{CLIP}^{\text{img}}$ is computed by encoding both the stroke renders and shape renders through a CLIP ViT/B-32 model, computing the cosine similarity, and scaling the score to [0-1]. Note that we only use the ResNet variants of CLIP for our optimization. 

\input{sec/figure_only}

\section{Ablations}

\paragraph{No intermediate CLIP layers.} As established in \citep{vinker2022clipasso}, the intermediate CLIP layers are essential for capturing the geometric structure of the target. Optimizing on only the fully-connected CLIP output results in abstractions that have some semantic correspondence with the target but the specific geometric features are noisy (\cref{fig:ablfc}). 

\paragraph{No Stage 1.} Stage 1 is essential for obtaining abstractions with visual volume. Without it the abstractions are flattened, so look reasonable from certain angles but not in others (\cref{fig:ablstage1}). 

\paragraph{Freestyle renders.} We ablate on Freestyle render supervision in \cref{supp:ablfreestyle}, instead running our method using opaque surface renders. The resulting abstraction is reasonable, but misses important geometric feature detail in the wheels and side mirrors of the car. 

\paragraph{SDF loss.} We ablate on the SDF loss in \cref{supp:ablsdf}. The SDF loss prevents texture features from floating off the surface implied by the rest of the strokes, such as the spots on Bob circled in red. 

\begin{figure}
    \centering
    \includegraphics[width=\linewidth]{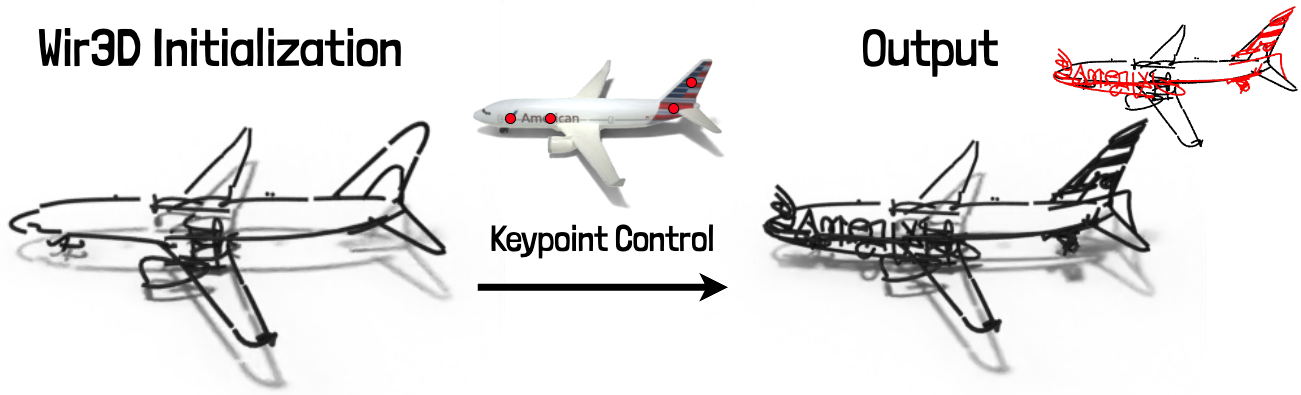}
    \caption{\textbf{Texture keypoint control.} We expand on the keypoint control example shown in the main paper with a textured example. We show how by selecting keypoints on the texture on the plane, we are able to refine the abstraction by incorporating those texture elements.}
    \label{supp:keypointcontrol}
\end{figure}


\section{Additional Applications}
\paragraph{Detail Refinement.} We show an additional example of keypoint-based abstraction refinement in \cref{supp:keypointcontrol}. For our refinement application, we freeze the existing curve set and optimize 6 new curves randomly initialized in a local Gaussian around each keypoint. We use the same losses as the main method and only sample views where the keypoints are visible, and optimize for 100 iterations. 

\paragraph{Curve-Based Shape Deformation.} Our deformation application exploits the close correspondence between the optimized curves and key visual features on the input surface, thanks to the SDF and keypoint localization losses. We develop a simple skinning system for the surface where each vertex is assigned a set of skinning weights to points sampled on all the curves in the scene. These skinning weights are based on the L2 distance between each vertex and sampled point, and a softmax is applied to ensure they sum to 1. Transformations to each curve can then be automatically mapped to the surface through these skinning weights, and the procedure can be performed at interactive speeds. We implement this deformation system as a proof-of-concept script, and show videos of the working system in the supplemental material, with screenshots displayed in \cref{fig:deformation}. Note that \emph{no smoothing postprocesses} are applied to the mapped transformations, and the smoothness of the deformations are a result of the effectiveness of the curves in interpolating the quantities along the surface.   

\begin{figure}
    \centering
    \includegraphics[width=\linewidth]{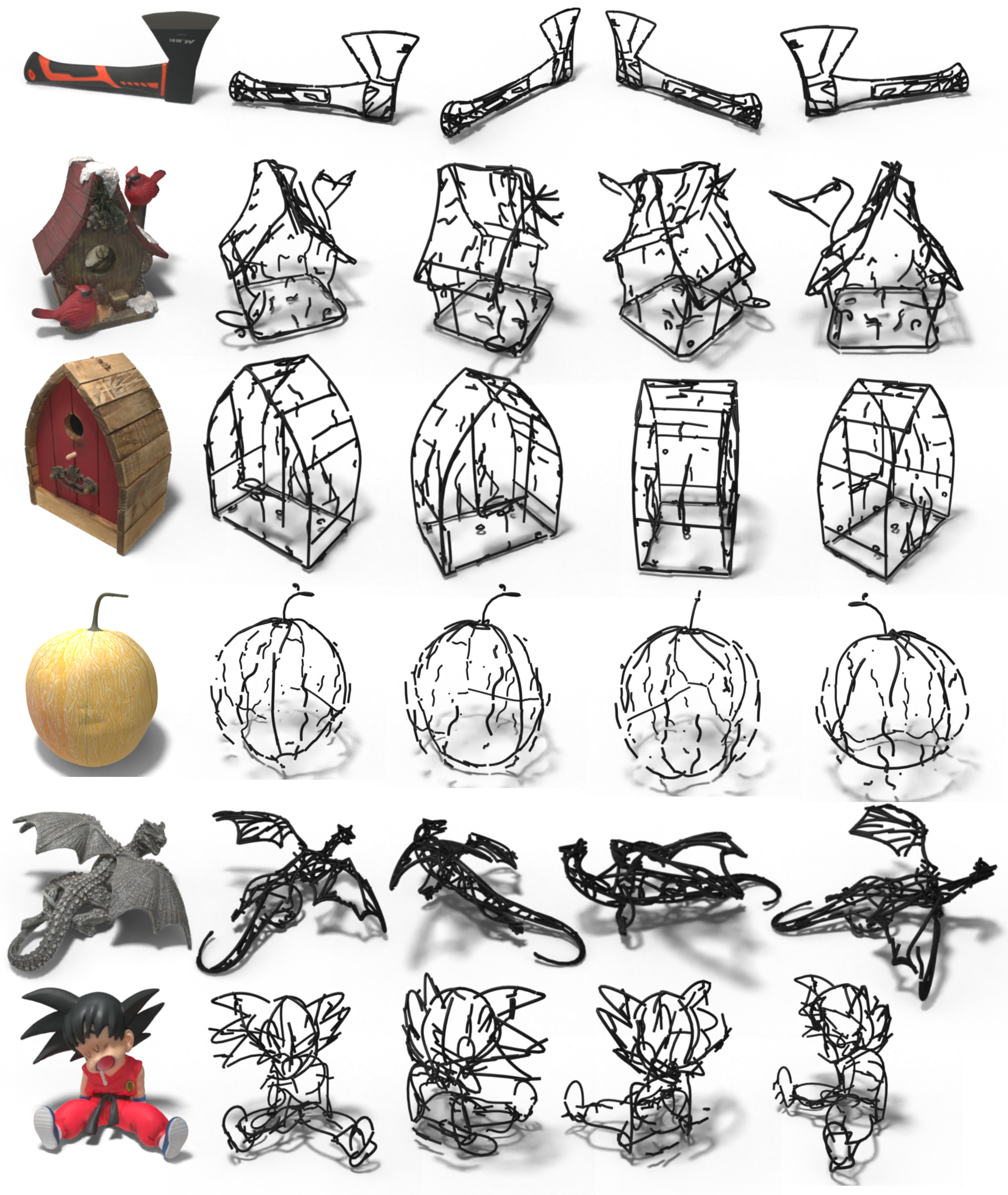}
    \caption{\textbf{Additional textured results.} We show additional textured results from the Meta DTC dataset \citep{pan2023aria}.}
    \label{supp:textured}
\end{figure}

\begin{figure*}
    \centering
    \includegraphics[width=\linewidth]{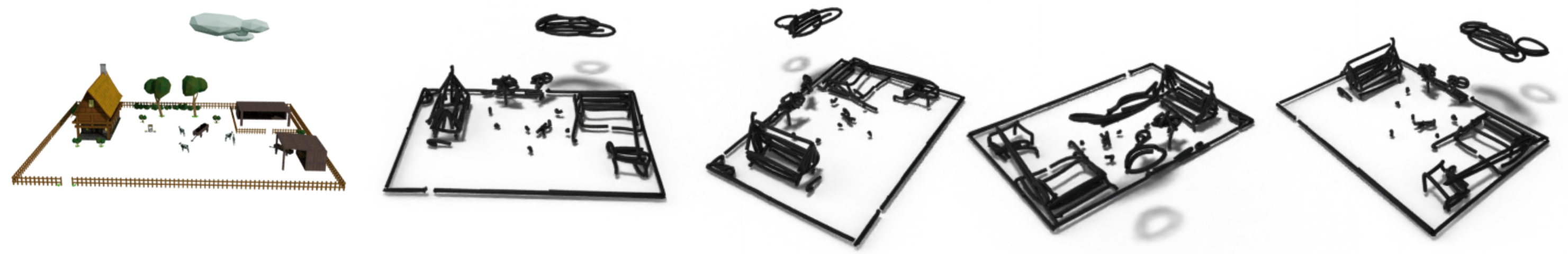}
    \caption{\textbf{Scene abstraction.} Our method extends to scene abstraction. Note how our method reproduces the global scene layout and captures all the objects in the scene despite the large scale differences.}
    \label{supp:scene}
\end{figure*}

\section{Additional Abstraction Results} 

\paragraph{Texture Abstractions.} Additional results on textured shapes are shown in \cref{supp:textured}. Our method is robust to many different types of models ranging from manufactured shapes with sharp edges to organic shapes with complex curvature. 

\paragraph{Scene Abstraction.} We show an example of our method run on a large scene in \cref{supp:scene}. Our method is able to reproduce the global scene layout, and successfully abstracts objects at different scales in the scene (e.g. house, trees, animals). 

\paragraph{Multi-View Fidelity.} Our curves are defined in 3D, so our abstraction is view-consistent by construction. However, this does not guarantee the curves plausibly represent the shape from arbitrary views. \cref{fig:fidelity} shows that our abstraction faithfully represents the shape for densely sampled views in a 360 range. 

\section{Optimization Details.} For both stages, we optimize for 20000 iterations, sample 1 view per iteration, and use an ADAM optimizer with a learning rate of 1e-3. For CLIP supervision we sample 4 augmentations per view. In stage 1 of the optimization, we use the RN101 CLIP architecture, with $\lambda_{\text{fc}} = 0.1$. In stage 2, we use the RN50x16 architecture, $\lambda_{\text{lpips}} = 0.1$, $\lambda_{\text{fc}} = 75$, and $\sigma=0.1$. 

\section{User Study Screenshots}
\begin{figure*}
    \centering
    \includegraphics[width=0.45\linewidth]{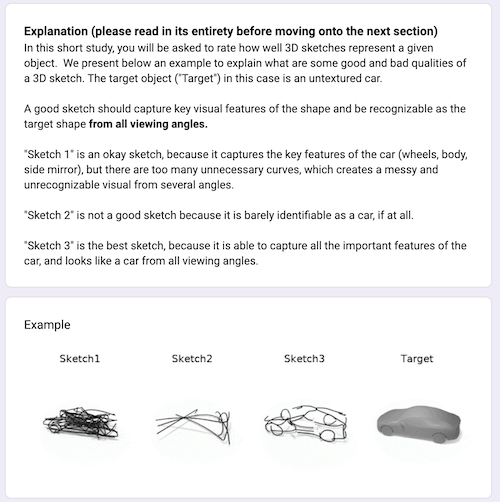}
    \includegraphics[width=0.45\linewidth]{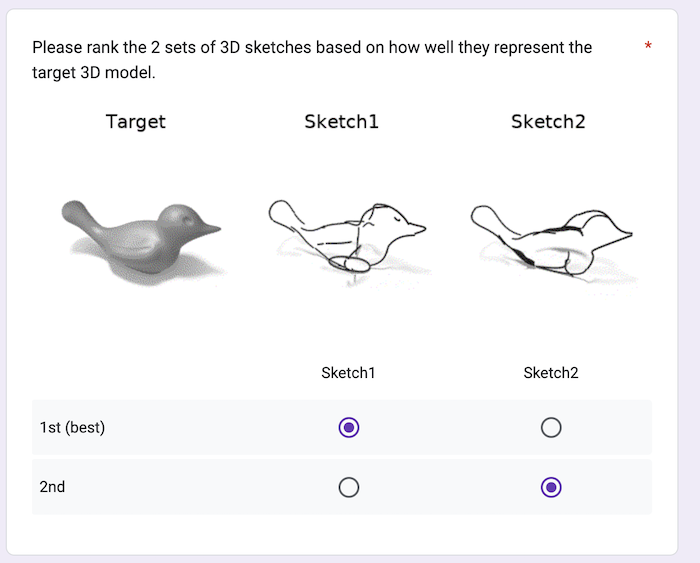}
    \includegraphics[width=0.45\linewidth]{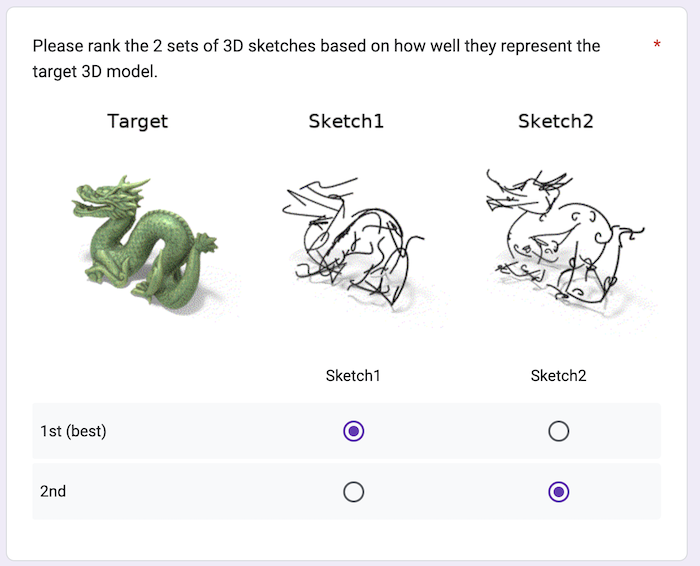}
    \includegraphics[width=0.45\linewidth]{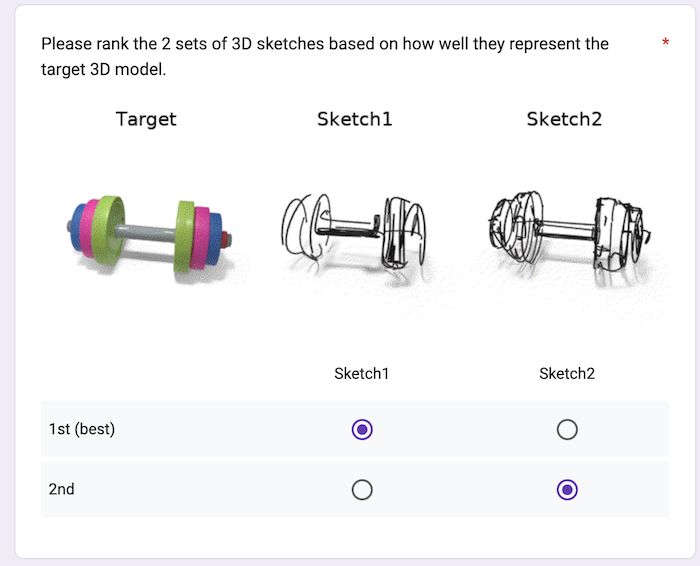}
    \caption{\textbf{Perceptual Study Screenshots.} Screenshots from our perceptual study. The question order and the order of Wir3D versus 3Doodle assignment to ``Sketch1/Sketch2" are randomized for each respondent.}
    \label{supp:userstudy}
\end{figure*}

We show screenshots from our user study in \cref{supp:userstudy}. The question order and the order of Wir3D versus 3Doodle assignment to ``Sketch1/Sketch2" are randomized for each respondent. All results shown are rotating gifs, so that users can evaluate based on the full 360 views of the abstraction. At the beginning of the study, we present three examples of abstractions of different quality and explain the factors that determine their quality, so that users can make more precise judgments in their visual evaluation.

\begin{figure*}
    \centering
    \includegraphics[width=0.45\linewidth]{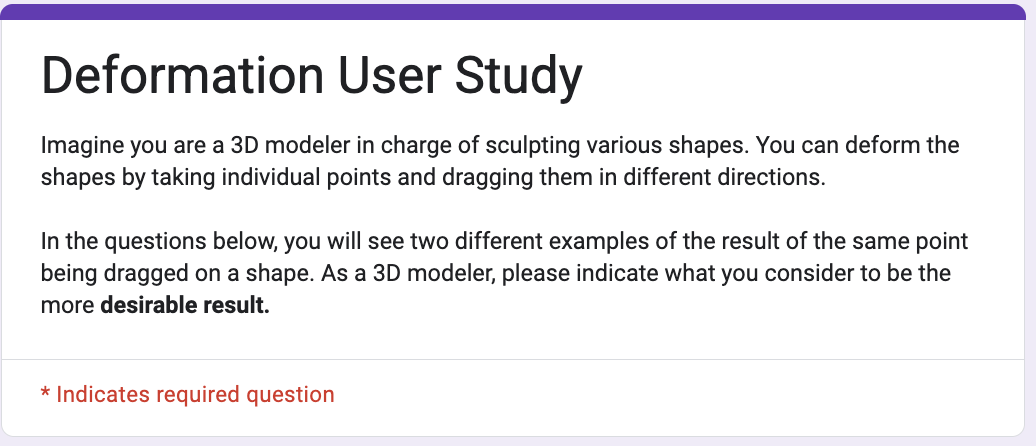}
    \includegraphics[width=0.45\linewidth]{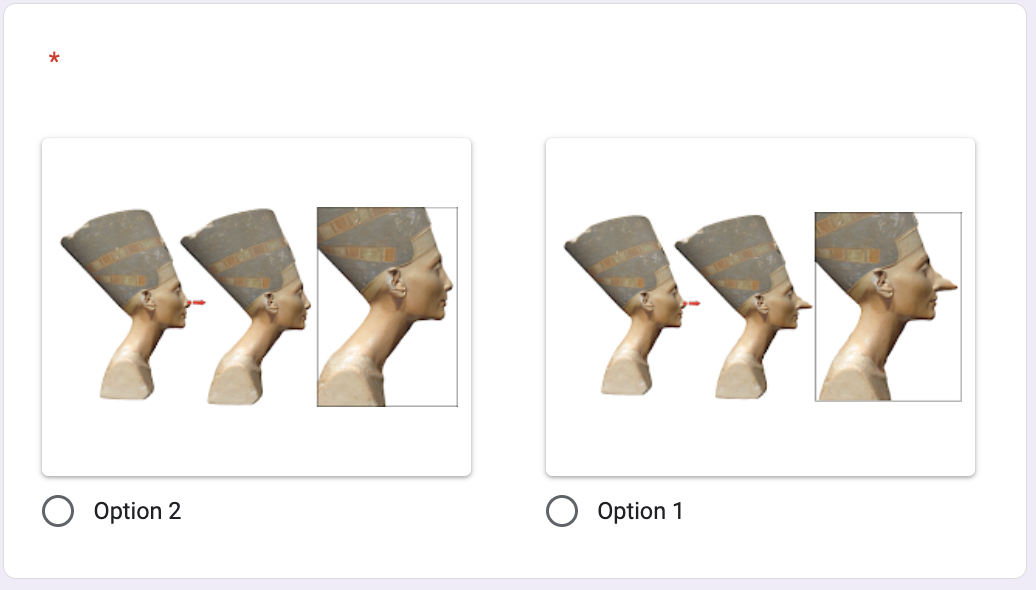}
    \includegraphics[width=0.45\linewidth]{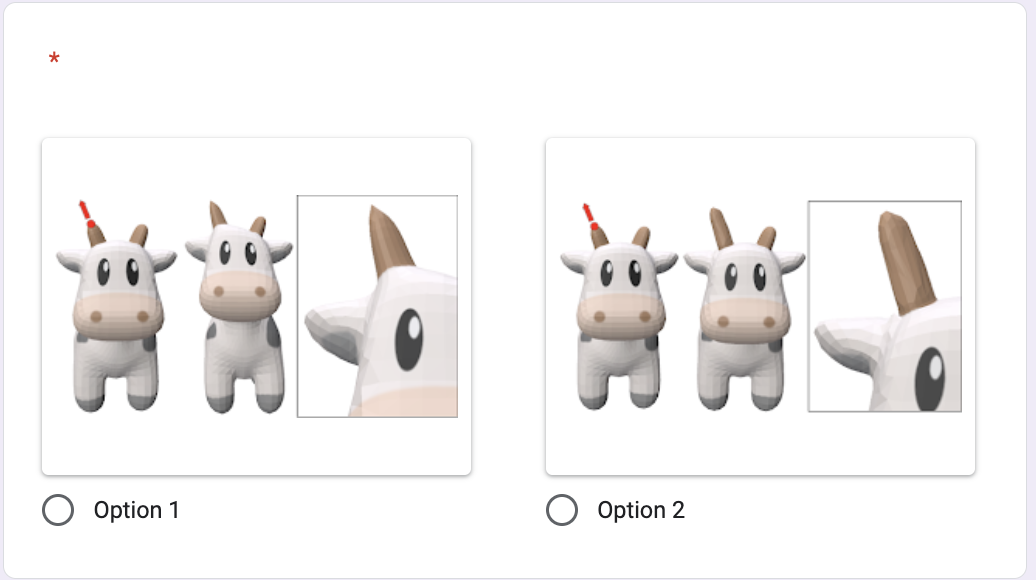}
    \includegraphics[width=0.45\linewidth]{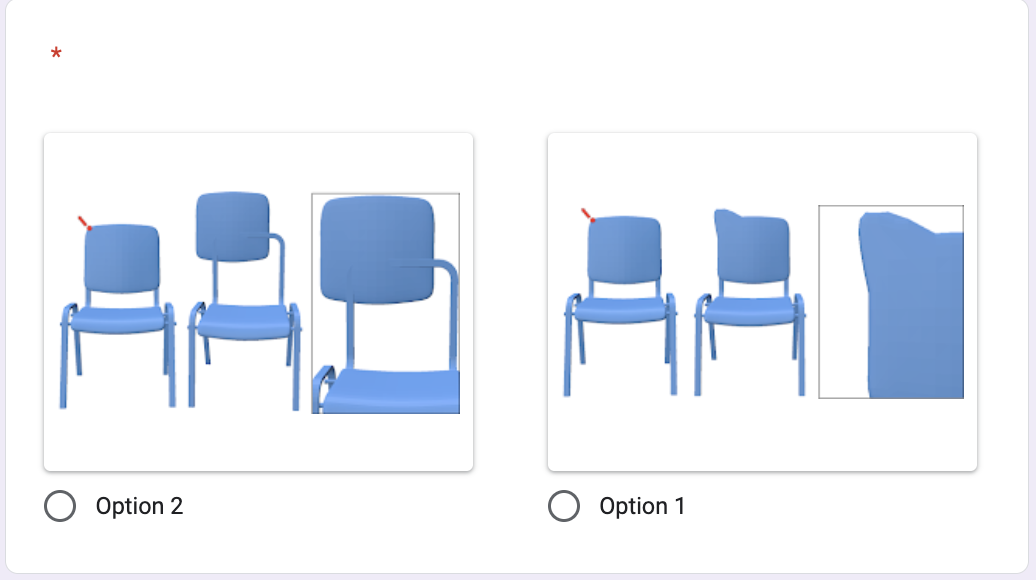}
    \caption{\textbf{Deformation Application User Study.} Screenshots from our user study comparing our deformation application using WIR3D curves as handles against ARAP \cite{sorkine_as-rigid-as-possible_2007} The question order and the order of Wir3D versus 3Doodle assignment to ``Option 1/Option 2" are randomized for each respondent.}
    \label{supp:deformation_userstudy}
\end{figure*}

\begin{figure*}
    \centering
    \includegraphics[width=\linewidth]{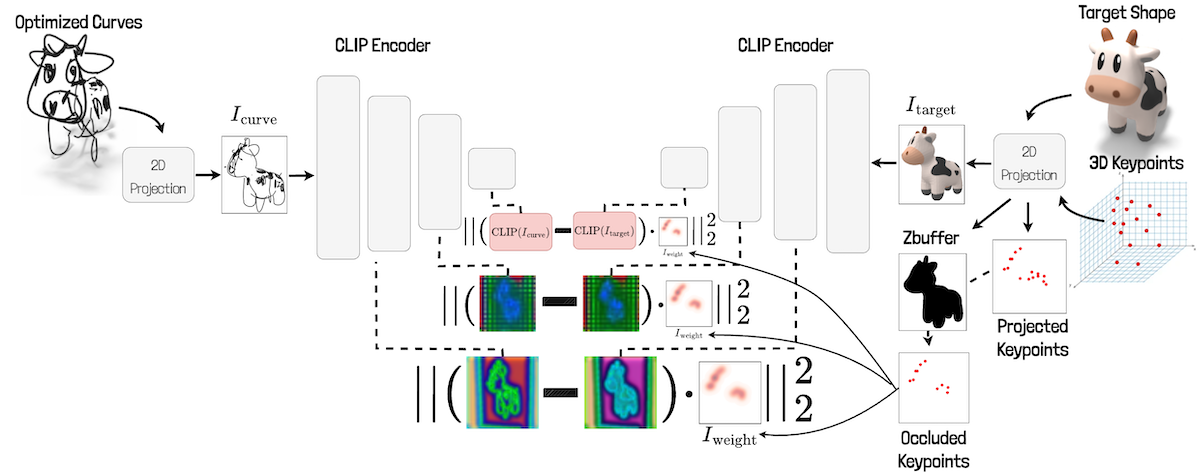}
    \caption{\textbf{Comprehensive Localized Keypoint Loss.} We show a comprehensive visualization of the localized keypoint loss. }
    \label{supp:spatial}
\end{figure*}

%% file: sec/figure_only.tex

\begin{figure}
    \centering
    \includegraphics[width=\linewidth]{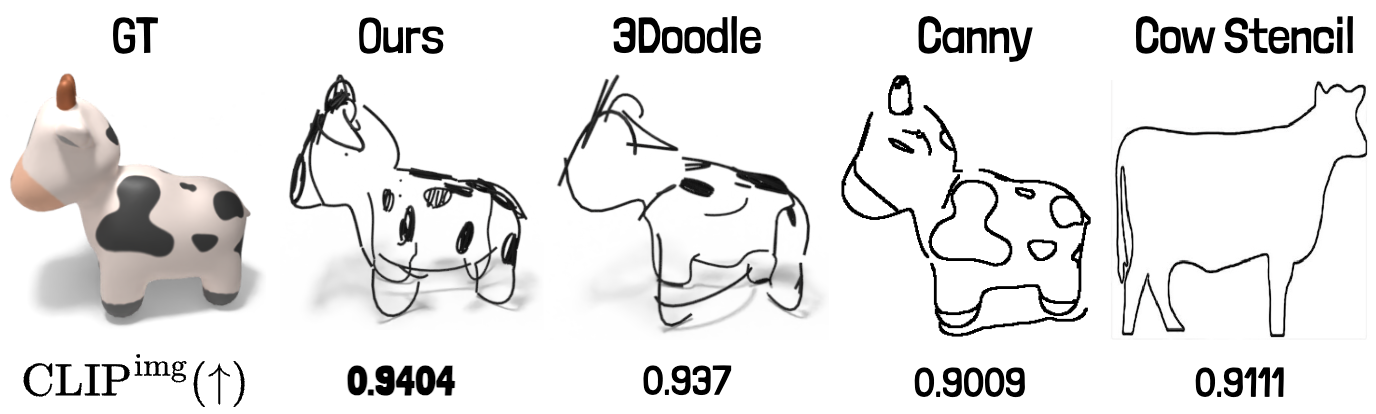}
    \caption{\textbf{Perceptual metrics reliability.} We show the unreliability of $\text{CLIP}^{\text{img}}$ in evaluating semantic similarity of curve abstraction to a target. We show for a given view, our stroke abstraction, 3Doodle's, the edge map for the view extracted using Canny edge detection \citep{canny_computational_1986}, and a random image of a cow stencil obtained from Google. Note that though the Canny edge map captures the entire geometric structure and textures of the shape, its $\text{CLIP}^{\text{img}}$ score is shockingly lower than that of the stencil image. The vast difference in the two images also demonstrates how small differences in score can indicate major differences in quality.}
    \label{fig:metricrobustness}
\end{figure}

\begin{figure}
    \centering
    \includegraphics[width=\linewidth]{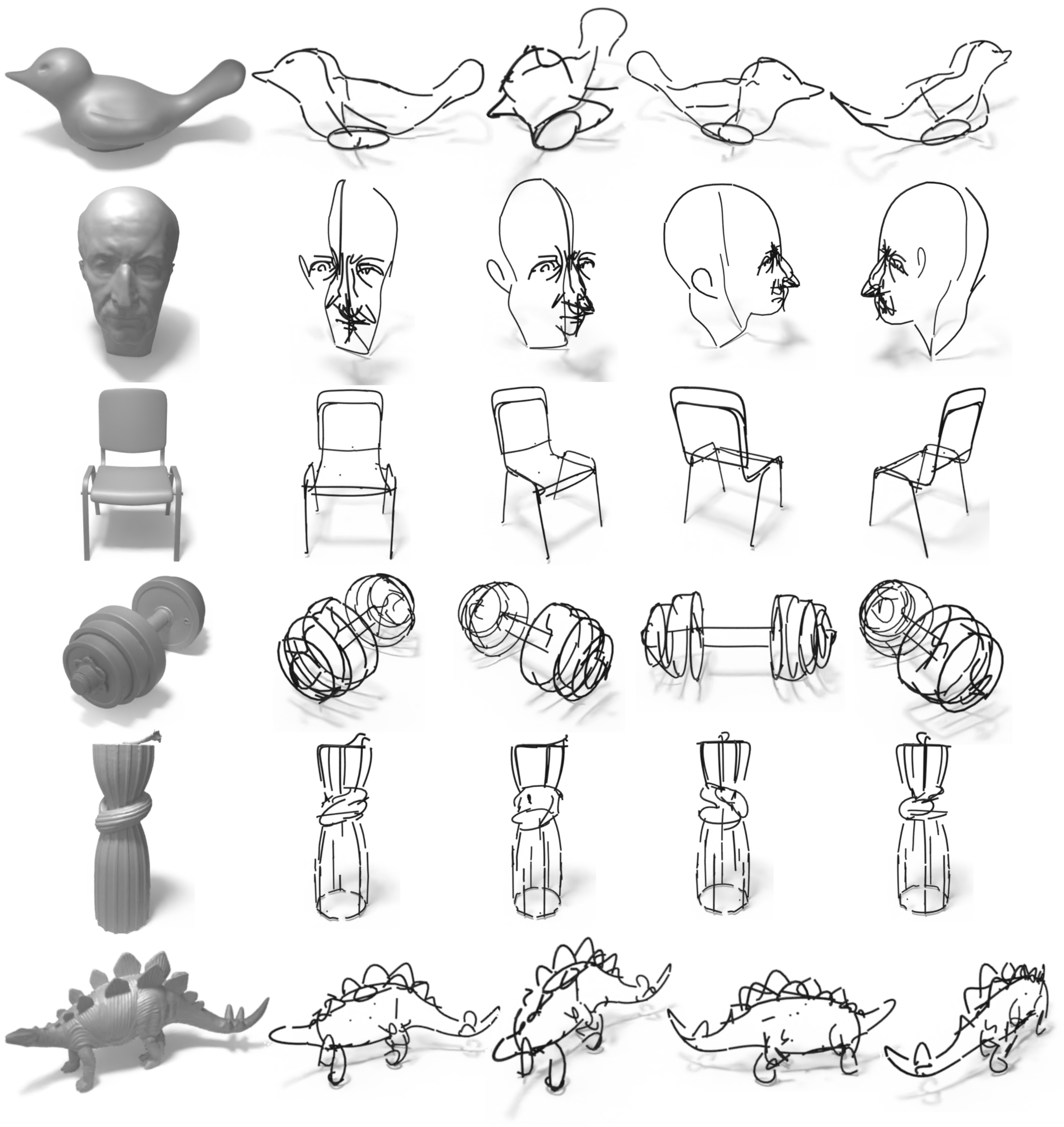}
    \caption{\textbf{Qualitative results for untextured shapes.} We show the result of our method on a collection of untextured meshes. Our method is effective and robust on a wide collection of different geometrie, such as the spiral column band (Row 6), the parallel rows of spines on the stegosaurus (Row 5).}
    \label{fig:untextured}
\end{figure}

\begin{figure*}[!t]
\centering
\includegraphics[width=0.99\linewidth, trim=0 0 0 0, clip]{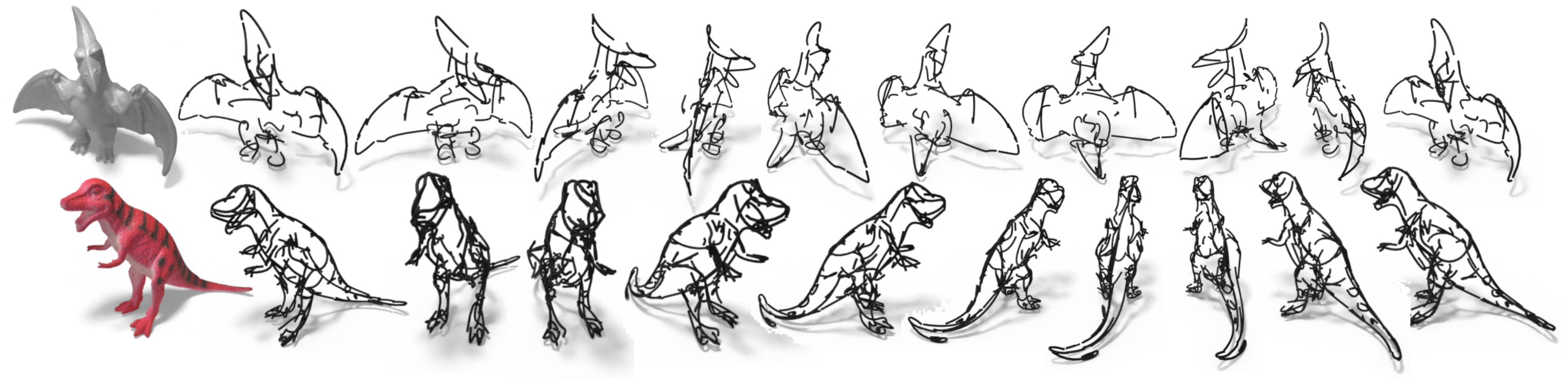}
\caption{\textbf{Multi-view fidelity.} \ourmethod{} adheres to the abstracted object in a 3D-consistent manner such that its properties can be perceived from every viewing angle.}
\label{fig:fidelity}
\end{figure*}

\begin{figure}[b]
    \centering
    \includegraphics[width=\linewidth]{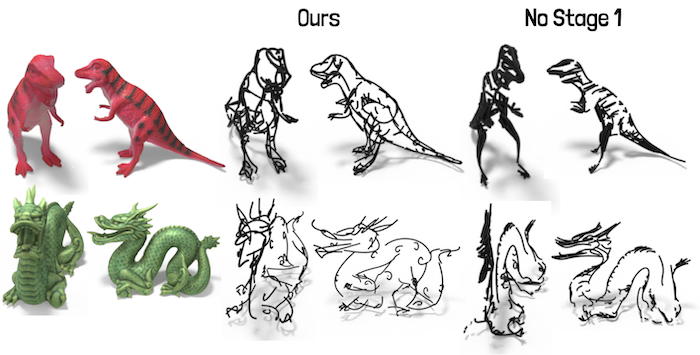}
    \caption{\textbf{Stage 1 ablation.} Stage 1 optimization is essential for capturing the full extent of the input geometry. Without it, the optimization tends to bias towards certain views, while the overall abstraction experiences a ``flattening" effect.}
    \label{fig:ablstage1}
\end{figure}

\begin{figure}[b]
    \centering
    \includegraphics[width=\linewidth]{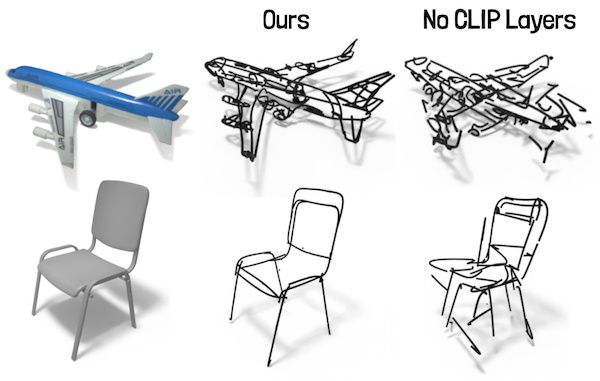}
    \caption{\textbf{CLIP layers ablation.} Supervising with the intermediate activations of CLIP is critical for maintaining coherent geometry. Using only the fully-connected CLIP output results in rough semantic abstraction, but the input shape geometric features are not well-preserved.}
    \label{fig:ablfc}
\end{figure}



\begin{figure}
    \centering
    \includegraphics[width=\linewidth]{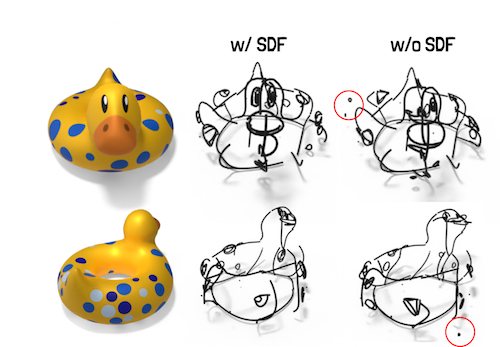}
    \caption{\textbf{SDF ablation.} The SDF loss helps to ensure abstracted visual features will stay anchored to the surface implied by the strokes. Without it, some features may hover outside the surface, such as the smaller spots on Bob.}
    \label{supp:ablsdf}
\end{figure}

